%% file: main.tex
\documentclass[acmsmall]{acmart}

\usepackage{enumitem}
\usepackage{subcaption}
\usepackage{multirow}
\usepackage{array}
\usepackage{diagbox}
\usepackage{xspace}
\usepackage{threeparttable}
\newcommand{\ie}{\emph{i.e.,}\xspace}
\newcommand{\eg}{\emph{e.g.,}\xspace}
\newcommand{\etal}{\emph{et al.}\xspace}
\newtheorem{finding} {Finding}

\AtBeginDocument{%
  }

\setcopyright{acmcopyright}
\copyrightyear{2022}
\acmYear{2022}
\acmDOI{XXXXXXX.XXXXXXX}

\acmJournal{JACM}
\acmVolume{37}
\acmNumber{4}
\acmArticle{1}
\acmMonth{8}

\begin{document}

\title{A Critical Study on Data Leakage in Recommender System Offline Evaluation}

\author{Yitong Ji}
\affiliation{%
  \institution{Nanyang Technological University}
  \country{Singapore}
}
\email{yitong.ji@ntu.edu.sg}

\author{Aixin Sun}
\affiliation{%
  \institution{Nanyang Technological University}
  \country{Singapore}
}
\email{axsun@ntu.edu.sg}

\author{Jie Zhang}
\affiliation{%
  \institution{Nanyang Technological University}
  \country{Singapore}
}
\email{zhangj@ntu.edu.sg}

\author{Chenliang Li}
\affiliation{%
  \institution{Wuhan University}
  \country{China}
}
\email{cllee@whu.edu.cn}

\begin{abstract}

Recommender models are hard to evaluate, particularly under offline setting. In this paper, we provide a comprehensive and critical analysis of the data leakage issue in recommender system offline evaluation. Data leakage is caused by not observing global timeline in evaluating recommenders \eg train/test data split does not follow global timeline. As a result, a model learns from the user-item interactions that are not expected to be available at prediction time. We first show the temporal dynamics of user-item interactions along global timeline,  then explain why data leakage exists for collaborative filtering models. Through carefully designed experiments, we show that all models indeed recommend future items that are not available at the time point of a test instance, as the result of data leakage. The experiments are conducted with four widely used baseline models - BPR, NeuMF, SASRec, and LightGCN, on four popular offline datasets - MovieLens-25M, Yelp, Amazon-music, and Amazon-electronic, adopting leave-last-one-out data split.\footnote{Our codes and datasets are available online \url{https://github.com/putatu/dataLeakageRec}.} We further show that data leakage does impact models' recommendation accuracy. Their relative performance orders thus become unpredictable with different amount of leaked future data in training.  To evaluate recommendation systems in a realistic manner in offline setting, we propose a timeline scheme, which calls for a revisit of the recommendation model design.
\end{abstract}

\setcopyright{acmcopyright}
\acmJournal{TOIS}
\acmYear{2022} \acmVolume{1} \acmNumber{1} \acmArticle{1} \acmMonth{1}\acmPrice{15.00}\acmDOI{10.1145/3569930}



\maketitle

\section{Introduction}
\label{sec:intro}

Recommender systems have gained significant attention from both academia and industry over the past decade. Many solutions have been proposed, from popularity to nearest neighbor based methods, then to model-based recommendation algorithms~\cite{recSysHandbook, rigorousEvaluation, recSurveyCF, recbole}. Come as no surprise, recent years have witnessed rapid advancement of deep learning based recommender models~\cite{recSysSurveyZhang}. Nevertheless, there are also questions on what real progress has been made in this research area~\cite{worrying}, and reports on model performance inconsistency due to different data splitting strategies and/or other factors~\cite{exploreDataSplit,AnelliBFMMPDN21-evaluation, futureItemRec}. These questions and reports call for a revisit on the evaluation of recommender system: \textit{are recommendation models evaluated in a realistic manner?} In fact, in our recent study, we show that in many studies, the simplest popularity baseline is not evaluated realistically~\cite{popularity20}.

Before we move on to performance evaluation issues, we revisit the recommender system problem definition. We refer to the recommender system handbook~\cite{recSysHandbook} and five survey papers~\cite{recSysHandbook, recSysSurveyZhang, biasDebiasSurvey, knowledgeGraphSurvey, evalRecSysSurveyFramework} for a generic problem definition of \textit{recommender system}: learning users' preferences from historical data, then to predict the item a user will rate/interact with in the nearer future. This definition is outlined on the basis of a \textbf{global timeline}, whereby ``history'' and ``future'' do not overlap along the timeline for any particular user at any time point. 

The aforementioned definition details a top-$N$ recommendation task, \ie recommending $N$ items that a user is most interested in. Note that, other than top-$N$ recommendation, rating prediction is also a widely studied task in recommender system~\cite{evalMetricsForTasks}. A top-$N$ recommendation task is fundamentally different from a rating prediction task in recommendation. The latter focuses on predicting the rating score that a user would give to an item, instead of predicting whether a user will rate an item. In this paper, we focus on top-$N$ recommendation task because top-$N$ recommendation task has been studied more extensively in recent years. Moreover, the ultimate goal of rating prediction is to find the items a user would give higher ratings and subsequently recommend the user with the corresponding items. In other words, a rating prediction task is to boost top-$N$ recommendation accuracy. Hence, we study top-$N$ recommendation in this paper.

\begin{table*}[t]
  \begin{center}
    \caption{Statistics of the four datasets used in this study}
    \label{tab:datasetStats}
    \begin{tabular}{l|ccrrrcc} 
      \toprule
    Dataset & Time span selected &  Data filtering & \#User & \#Item & \#Interaction \\
      \midrule
      MovieLens-25M & 21/11/2009 to 20/11/2019 & No filtering & $62,202$ & $56,774$ & $9,808,925$ \\
      Yelp & 13/12/2009 to 12/12/2019 & 10-core & $116,655$ & $61,027$ & $3,127,215$ \\
      Amazon-music & 02/10/2008 to 01/10/2018 & 5-core & $11,651$ & $9,243$ & $114,833$  \\
      Amazon-electronic & 05/10/2008 to 04/10/2018 & 10-core & $109,990$ & $39,552$ & $1,752,238$ \\
      \bottomrule
    \end{tabular}
  \end{center}
\end{table*}

We argue that although \textbf{global timeline} is specified in the aforementioned recommender system problem definition, it is not observed in many offline evaluation settings for recommender system. By ``not observing global timeline'', we are referring to the case whereby events are not considered in chronological order along the global timeline. We argue that ignorance of the timeline leads to two major issues.

\begin{figure*}[t]
    \centering
	  \begin{subfigure}[t]{0.35\columnwidth}
    	\centering
    	\includegraphics[width=\columnwidth]{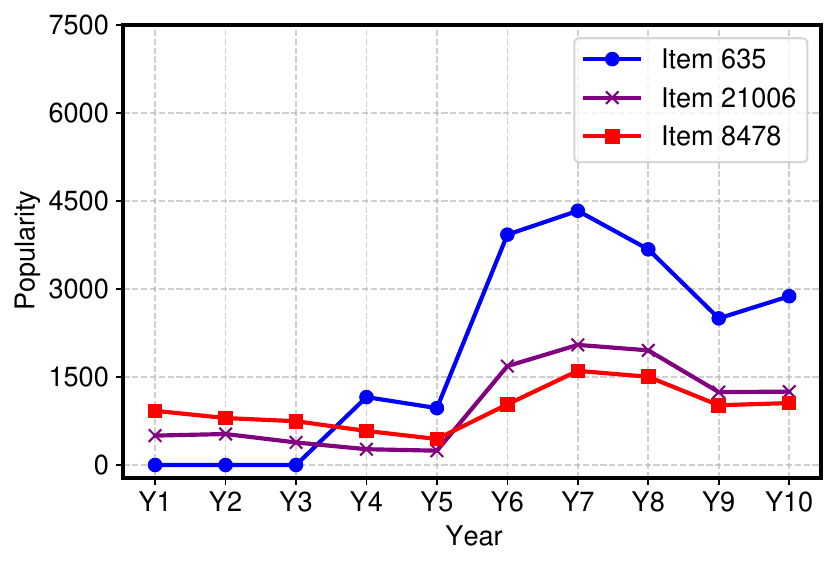}
    	\caption{MovieLens-25M}
        \label{sfig:movielens_pop}
	\Description{}
    \end{subfigure}
    \quad
	  \begin{subfigure}[t]{0.35\columnwidth}
    	\centering
    	\includegraphics[ width=\columnwidth]{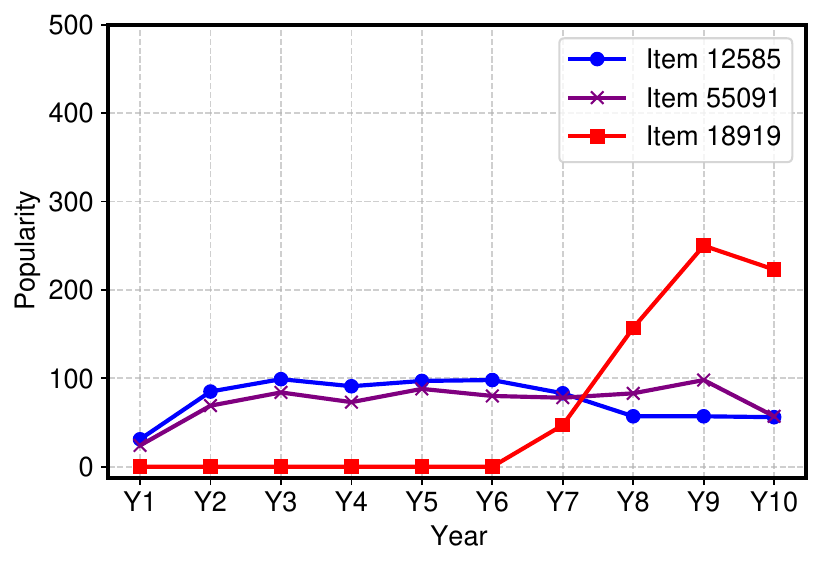}
    	\caption{Yelp}
        \label{sfig:yelp_pop}
	\Description{}
    \end{subfigure}
    \caption{Popularity of three sampled items from two datasets, over ten-year period ($Y1$ to $Y10$).}
    \label{fig:pop_globaltimeline}
\end{figure*}

\begin{figure*}[t]
    \centering
	  \begin{subfigure}[t]{0.4\columnwidth}
    	\centering
    	\includegraphics[width=\columnwidth]{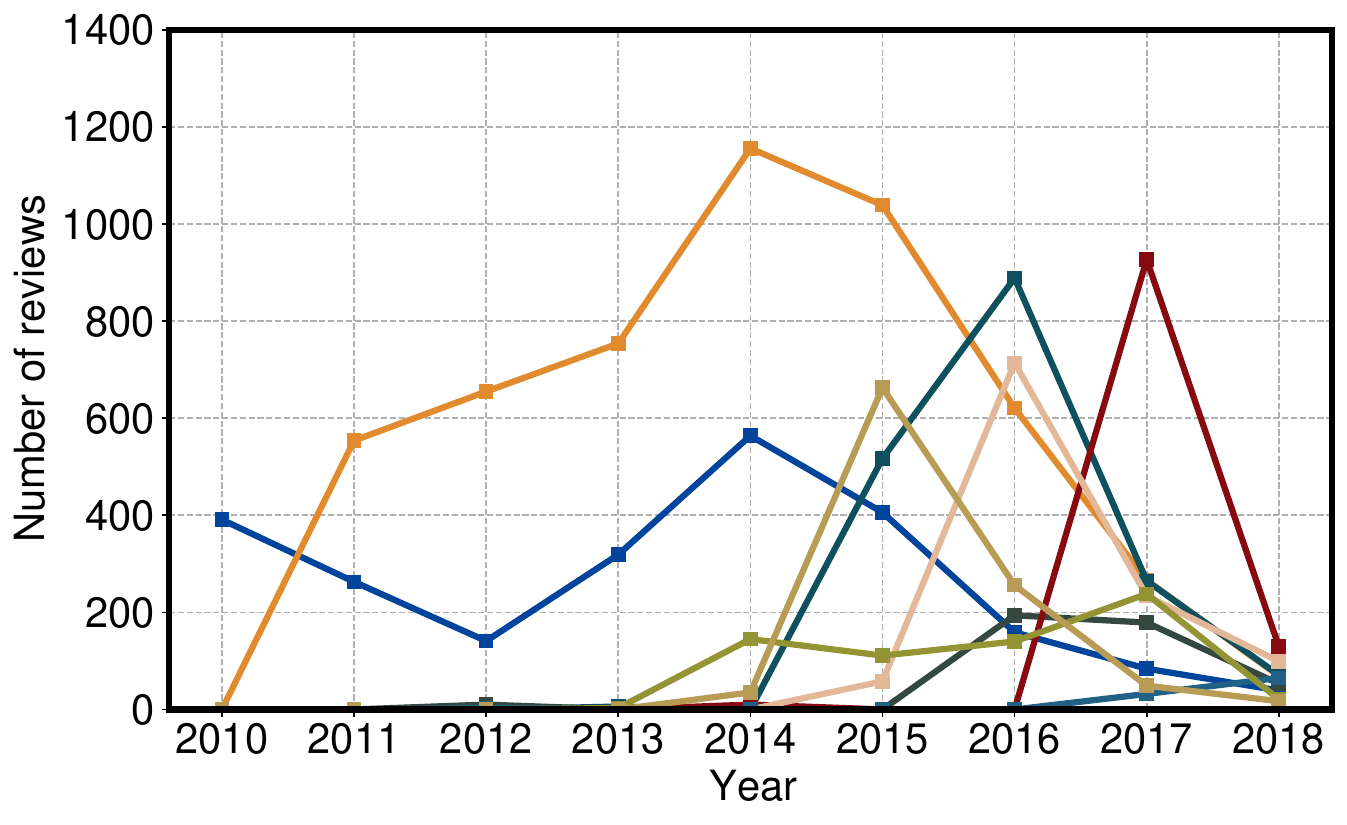}
    	\caption{Activeness of Ipad models}
        \label{sfig:ipad_pop}
	\Description{}
    \end{subfigure}
    \quad
	  \begin{subfigure}[t]{0.4\columnwidth}
    	\centering
    	\includegraphics[ width=\columnwidth]{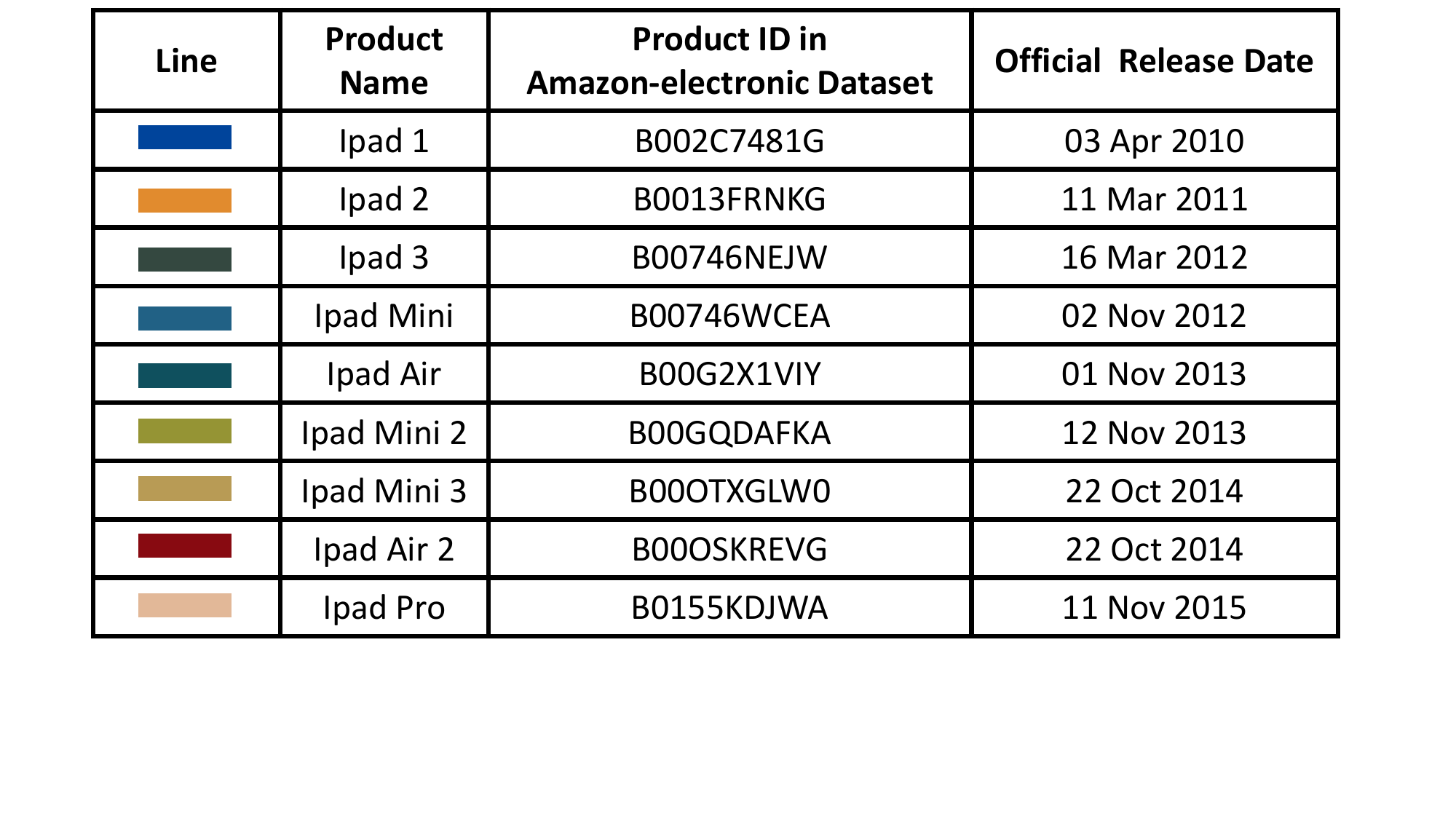}
    	\caption{Ipad models' official release dates and their IDs in Amazon-electronic dataset}
        \label{sfig:ipad_info}
	\Description{}
    \end{subfigure}
    \caption{Activeness of Ipad models from Year 2010 to Year 2018, in Amazon-electronic dataset. The table lists the release date of different Ipad models and their corresponding product ID in Amazon-electronic dataset.}
    \label{fig:ipad_pop}
\end{figure*}

The first issue is that the models designed cannot capture the global temporal context of the user-item interactions. Temporal context along the timeline in datasets (\eg Douban~\cite{douban}, Amazon~\cite{amazon}) have been studied in~\cite{causalPopularityBias, timeMatters}. In this paper, we further introduce three examples to illustrate the temporal context in MovieLens-25M, Yelp and Amazon-electronic datasets. In Figure~\ref{fig:pop_globaltimeline}, we sample three items from two datasets: MovieLens-25M and Yelp (see Table~\ref{tab:datasetStats} for the details of the datasets). Each dataset contains ten-year user-item interactions. We plot the number of interactions the sampled items received in each year (\ie items' yearly popularity) over the ten years. Naturally, the popularity of items changes over time. In fact, not all items are available for  ratings/interactions from the beginning of the timeline (Year 1). Item $18919$ in Yelp receives its very first rating after Year 7 (the seventh year since the beginning of the ten-year time period) and becomes popular from then. We may consider that this item was first introduced to the review system in Year 7. However, most recommender models (including popularity-based recommenders) in the literature do not consider this temporal context in evaluation. Another illustrative example could be the releases of the Ipad models by Apple. In Figure~\ref{sfig:ipad_info}, we list nine Ipad models and their corresponding official release dates. It can be seen that, from 2010 to 2015, new Ipad models have been gradually introduced to the market. We then match the Ipad models with the products in the Amazon-electronic dataset; the corresponding product IDs of different Ipad models in the Amazon-electronic dataset can be found in Figure~\ref{sfig:ipad_info}. Then we plot the activeness (\ie number of reviews) of the Ipad models in the Amazon-electronic dataset, in Figure~\ref{sfig:ipad_pop}. From the plot, we make two observations: (1) The activeness of the Ipads in Amazon-electronic can reflect their release dates because the Ipads only start receiving reviews from their release years; (2) Generally, an Ipad model is more popular near its release year but slowly losing attentions from consumers after a few years. Taking the timeline from 2010 to 2018 as an example, if \textit{global timeline} is not observed, trends like newly released products and dynamic popularity of a product cannot be well captured.

The second issue is data leakage. In many evaluations, train/test data split does not follow global timeline (see Table~\ref{tab:dataPartitioning} for the commonly used data split strategies). The training set may contain interactions that happen after the test instances in the test set. Hence, a model learns from future user-item interactions to predict current user preference, due to collaborative filtering. We will explain the reason in Section~\ref{sec:data_leakage}. In reality, a model can never access data instances that happen in future. For example, a model shall not learn from user interactions with Ipad 3 that was released in 2012, to predict user's preference on Ipad 1 any time in the year of 2010. Learning from interactions that happen in future contradicts to the problem definition of recommender system. Thus, results obtained in such evaluation setting may not well reflect a model's true performance in online setting. 

\begin{table*}[t]
    \begin{center}
        \caption{Commonly used data split strategies in offline evaluation of recommender systems. We indicate whether local timeline (\eg time specific to a user)  or global timeline is observed, and whether a split strategy leads to data leakage.}
        \label{tab:dataPartitioning}
        \begin{tabular}{l|m{5.5cm}|ccc}
            \toprule
             Data split strategy & Definition of training and test instances  & Local & Global & Data\\
              & & timeline & timeline & leakage \\
             \midrule
             Random-split-by-ratio & Randomly sample a percentage of user-item interactions as test instances; the remaining are training instances. & No  & No & Yes \\
             \hline
             Random-split-by-user & Randomly sample a percentage of users, and take all their interactions as test instances; the remaining instances from other users are training instances. & No & No & Yes\\
            \midrule 
             Leave-one-out-split & Take each user's last interaction as a test instance; all remaining interactions are training instances. & Yes & No & Yes \\
             \hline
            Split-by-timepoint & All interactions after a time point are test instances; interactions before this time point are training instances. & No  & Yes & No\\
            \bottomrule
        \end{tabular}
        
    \end{center}

\end{table*}

In this paper, we conduct a critical analysis on the impact of data leakage in evaluating recommender system. Note that the notion of data leakage in recommender system was mentioned in earlier studies~\cite{predictFutureTaste,timeAware, seqHypergraph}. However, to the best of our knowledge, we are the first to provide a comprehensive explanation on the reason of data leakage, \ie collaborative filtering not observing global timeline in offline evaluation (see Section~\ref{sec:data_leakage}). Another main contribution of this paper is that, through carefully designed experiments, we quantify the impact of data leakage. The critical analysis of data leakage is done with four popular and strong models on four widely used datasets:  MovieLens-25M, Yelp, Amazon-music, and Amazon-electronic, adopting leave-one-out data split. 
The four models, Bayesian Personalized Ranking (BPR)~\cite{BPR}, Neural Matrix Factorization (NeuMF)~\cite{He2017}, SASRec~\cite{SASRec} and LightGCN~\cite{lightgcn}, are selected for (i) each representing a different modeling technique, and (ii) all four models are often used as baselines in literature. We summarize the findings from our experiments as follows: 
\begin{itemize}
  
    \item All four baseline models recommend ``future items'' to a past test instance. Here, future items are the items that are only available in the system after the time point of the test instance. For example, for a test instance occurred  some time in 2013, a model recommends Ipad Pro which is released two years later in 2015.  Recommending future items is a strong evidence that the mainstream offline evaluation setting, which ignores the global timeline, is invalid.
    \item ``Future training instances'' affect user-item interaction distributions. The training set with ``future data'' is not reflective of the temporal context at the time point when a test instance occurs. Hence, the top-$N$ recommendation lists from experiments with future data differ from those with no data leakage.
    \item Data leakage does affect a model's recommendation accuracy. The impact of data leakage on recommender models is \textit{unpredictable}. It is not true that more future data in training leads to higher accuracy. More future data may improve or deteriorate recommendation accuracy.
    \item The ranking orders of the baseline models vary with the inclusion of different amounts of ``future data'' in our experiments. Hence, it is impractical and meaningless to compare recommendation models' performance when data leakage exists in offline evaluation. We believe that not observing global timeline in evaluation is a major reason that reproducibility in recommender system is hard.  
\end{itemize}

Based on our understanding of recommendation system and what we have learned from this critical analysis, we propose a new \textit{timeline scheme} for recommender system offline evaluation in Section~\ref{sec:timelineScheme}. The essence of the timeline scheme evaluation is to observe global timeline in evaluation, which requires recommender models to learn from only historical interactions. We believe that the timeline scheme facilitates a more realistic evaluation of recommender systems conducted in an offline setting. After reviewing related studies in Section~\ref{sec:related}, we conclude this paper in Section~\ref{sec:conclude}.

\section{Data Leakage in Offline Evaluation}
\label{sec:data_leakage}
If a model is deployed in production, \ie an online setting, a global timeline will be naturally followed. The recommender can only learn from past user-item interactions and make online recommendations upon request. However, due to limited access to online platforms, recommender systems in academic research are often restricted to offline evaluation with static datasets~\cite{evalRecSysSurveyFramework, researchPaperRecSysOffline}. 

In a typical offline evaluation setting, given $m$ users and $n$ items in a dataset, a $m\times n$ user-item interaction matrix is constructed, where each entry in this matrix indicates the corresponding user-item interaction. The interaction can be explicit \eg a score (1 to 5) given by a user to an item, or be implicit \eg indicating whether the user purchases/rates/clicks this item (1 for purchase/rating/click and 0 otherwise). In this paper, \textit{user-item interactions are all implicit}. That is, if a user rates an item, we concern the action of rating, not the absolute value of the rating. Thus, we focus on the top-$N$ recommendation task, which is to predict the ``$1$'' entries in the user-item interaction matrix.

To compute recommendation accuracy, a subset of user-item interactions in this $m\times n$ matrix is masked as a test set; the remaining known entries are treated as the training set. A recommendation model is trained using the training set, then evaluated using the test instances. Listed in Table~\ref{tab:dataPartitioning}, many strategies have been used to sample the user-item interactions to be masked to form a test set~\cite{rigorousEvaluation, exploreDataSplit}. We note that only split-by-timepoint strategy observes global timeline. 

In most cases, the entire training set is fed to a recommender as a whole, to learn the recommendation model~\cite{rigorousEvaluation, offlineEvaluation}. That means, the training set is treated as a \textbf{static set}. Although some time-aware recommendation models~\cite{timeAware, SASRec, seqHypergraph, NARM} do include timestamp information of training instances as an attribute, they remain taking the entire training set as a \textit{static set}. These models may consider local timeline specific to an individual user or item, but are not designed to observe the global timeline. As a result, these models are inevitably evaluated with data leakage.

\subsection{Global Timeline and Local Timeline}
In our discussion, we emphasize the importance of considering global timeline. To support our discussion, we use four large datasets to illustrate the distribution of user-item interactions along the global timeline. The four datasets are MovieLens-25M\footnote{https://grouplens.org/datasets/movielens/25m/}, Yelp\footnote{https://www.kaggle.com/yelp-dataset/yelp-dataset}, Amazon-music\footnote{https://jmcauley.ucsd.edu/data/amazon/}, and Amazon-electronic\footnote{https://jmcauley.ucsd.edu/data/amazon/}. 
From each dataset, we extract interactions within a 10-year period and the statistics of the datasets are reported in Table~\ref{tab:datasetStats}. Details about the dataset preparation and data filtering (\eg $k$-core) techniques are reported in Section~\ref{ssec:dataset}.

Naturally, not all users are active for the entire 10-year period, and not every item is available to be rated (or be interacted with) over the entire 10 years. For instance, a user may register a Yelp account in 2018 and start to review businesses on Yelp from 2018; a movie that was released in 2018 can only receive ratings since then. We define the \textit{active time period} of a user to be the time duration between the user's first and  last interactions that are captured in a dataset. Similarly, \textit{active time period} of an item is the time duration between its first and last interactions received from any user in a dataset. From the first to the last interaction, the active time period is the \textit{local timeline} specific to any particular user or item.

\begin{figure*}[t]
	\centering
	  \begin{subfigure}[t]{0.4\columnwidth}
    	\centering
    	\includegraphics[width=\columnwidth]{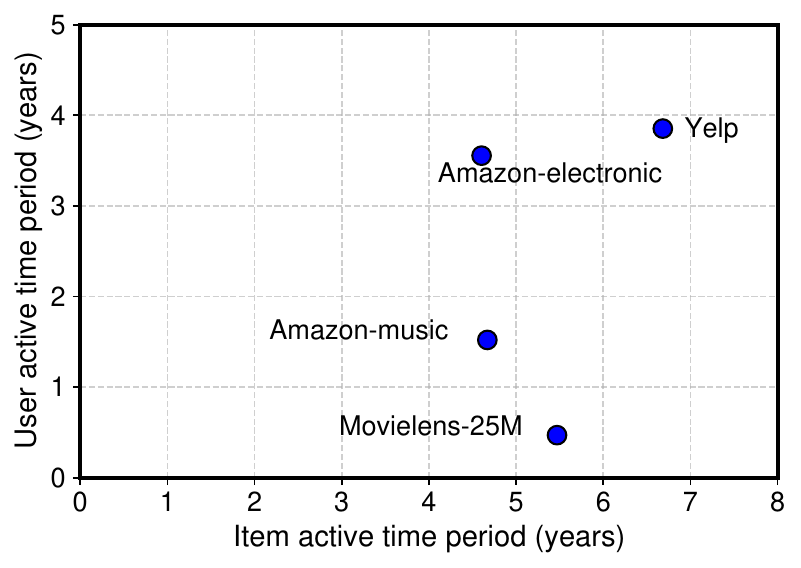}
    	\caption{Mean active time period in years.}
        \label{sfig:activeTimeMean}
	\Description{}
    \end{subfigure}
    \quad
	  \begin{subfigure}[t]{0.4\columnwidth}
    	\centering
    	\includegraphics[width=\columnwidth]{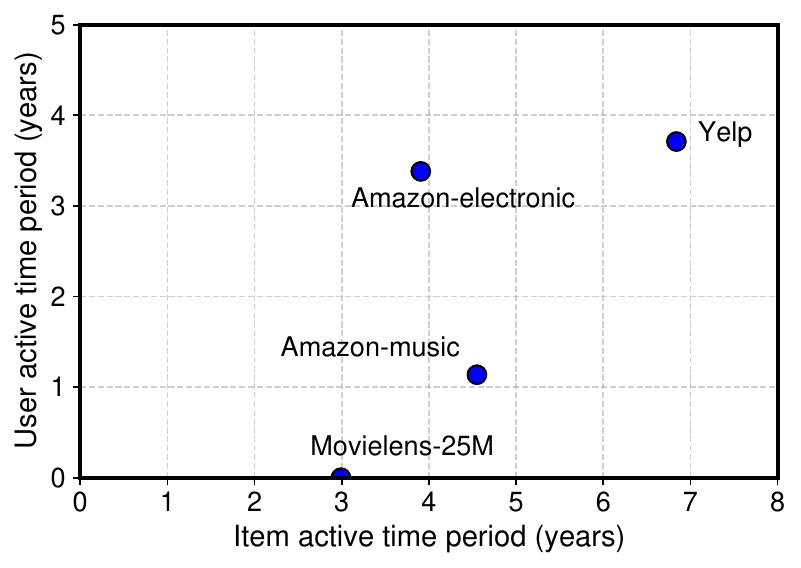}
    	\caption{Median active time period  in years}
        \label{sfig:activeTimeMedian}
	\Description{}
    \end{subfigure}
    \caption{Mean and median active time period of users and items in the four datasets.}
    \label{fig:activeTime}
\end{figure*}

Figure~\ref{fig:activeTime} provides a holistic view of the mean and median \textit{active time periods} of users and items, derived from the four datasets. Because the items (\eg movie, music, and restaurants) are different in nature, the four datasets show different distributions in terms of user and item activeness. Nevertheless, in all four datasets, mean/median active time of items are much longer than that of users. Specifically, the mean/median user active time periods in all datasets are shorter than 4 years, which is less than half of the 10-year period covered by the datasets. For MovieLens-25M, users tend to be active for a very short time period, shorter than 1 year. In fact, for a large number of users in Movielens dataset, all ratings of a user are recorded within a single day.\footnote{We note that MovieLens does not provide a reliable timestamp because many users are only active for one day. A similar observation is made in~\cite{pseudoSequence}. Although MovieLens is not a good dataset for this study, we include it due to its extreme popularity in academic research of recommender system~\cite{rigorousEvaluation, datasetDillema}.}  

\begin{figure*}[t]
	\centering
	  \begin{subfigure}[t]{0.45\columnwidth}
    	\centering
    	\includegraphics[width=\columnwidth]{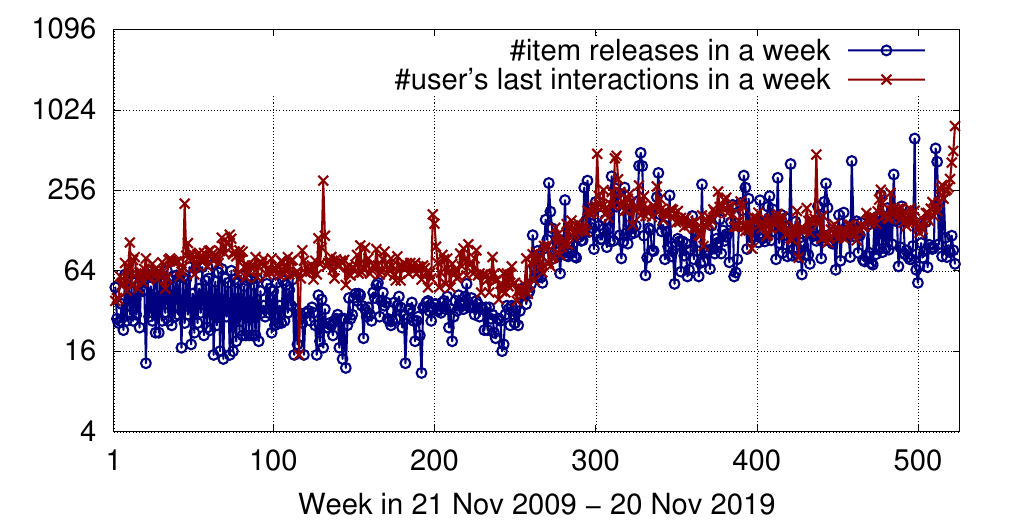}
    	\caption{MovieLens-25M}
        \label{sfig:movieLens_weekly_activity_with_release_log}
	\Description{}
    \end{subfigure}
    \quad
	  \begin{subfigure}[t]{0.45\columnwidth}
    	\centering
    	\includegraphics[width=\columnwidth]{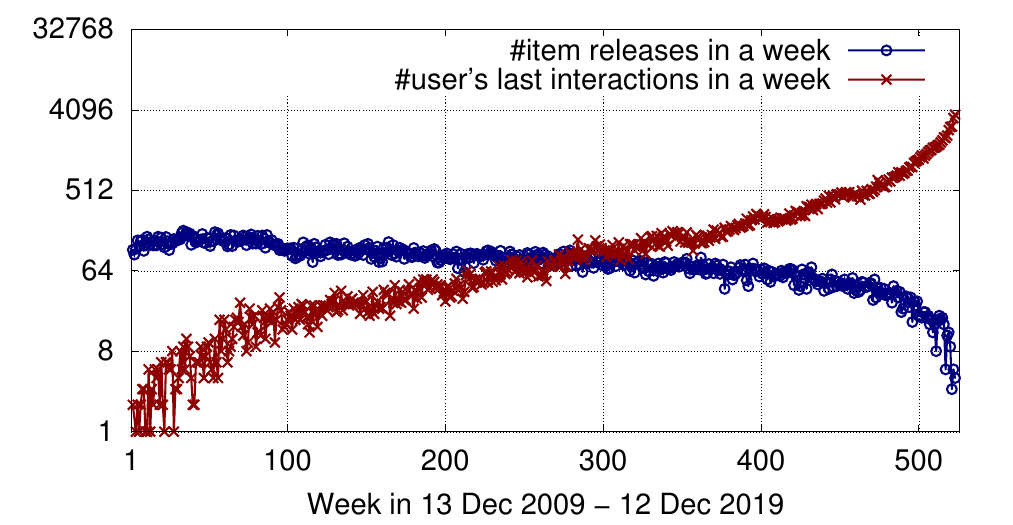}
    	\caption{Yelp}
        \label{sfig:yelp_weekly_activity_with_release_log}
	\Description{}
    \end{subfigure}
    \quad
    \begin{subfigure}[t]{0.45\columnwidth}
    	\centering
    	\includegraphics[width=\columnwidth]{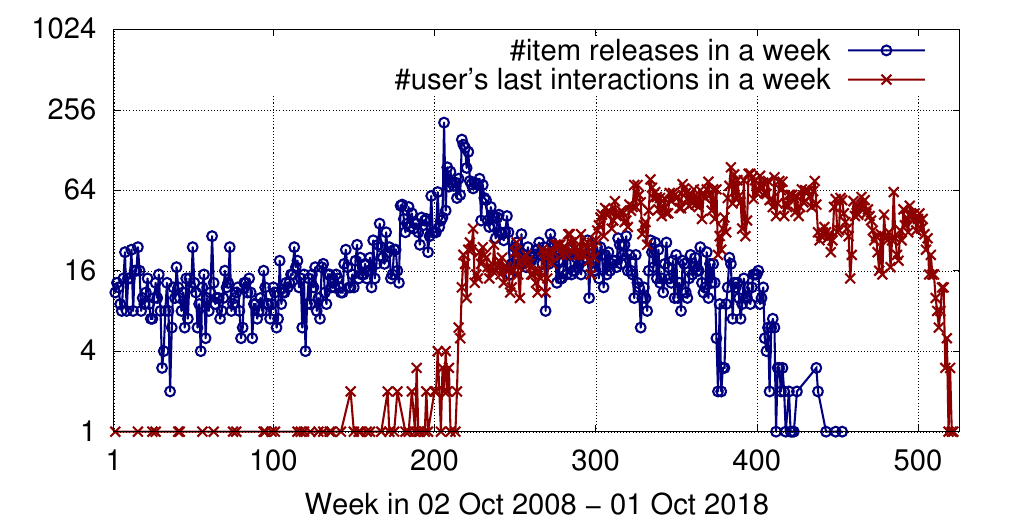}
    	\caption{Amazon-music}
        \label{sfig:amazonm_weekly_activity_with_release_log}
	\Description{}
    \end{subfigure}
    \quad
    \begin{subfigure}[t]{0.45\columnwidth}
    	\centering
    	\includegraphics[width=\columnwidth]{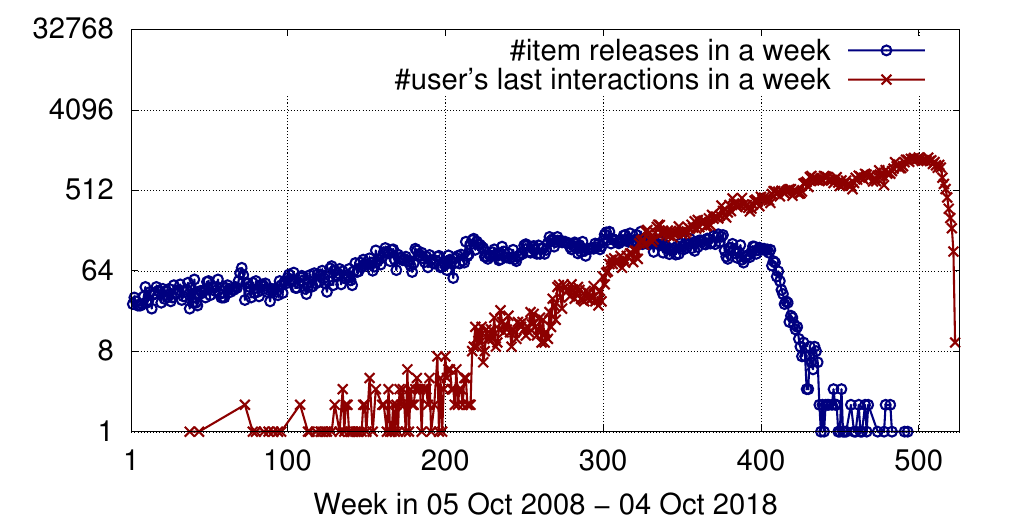}
    	\caption{Amazon-electronic}
        \label{sfig:amazone_weekly_activity_with_release_log}
	\Description{}
    \end{subfigure}
    \Description{}
    \caption{Users' last interactions and item releases in each week, in the 10-year period. (Best viewed in color)}
    \label{fig:weekly_activity_with_release_log}
\end{figure*}

To further illustrate the distribution of user/item activeness along the global timeline, Figure~\ref{fig:weekly_activity_with_release_log} plots the number of item releases and the number of users' last interactions occurred in each week.  We consider the time point at which an item receives its very first rating from any user its release time. The blue line in Figure~\ref{fig:weekly_activity_with_release_log} plots the number of items released in each week in the 10-year period, or 520 weeks. The other line plots the number of users whose \textit{last interactions} are recorded in each week over the 10-year period. Observe that in all four datasets, new items become available at any time, and users' last interactions occur at any time. 

In summary, not all users are active throughout the duration covered by a dataset, and not all items are available to be interacted with since the beginning of a dataset along the global timeline. In addition, as shown in Figure~\ref{fig:pop_globaltimeline}, along the global timeline, user-item interactions demonstrate temporal dynamics. The same observation of temporal dynamics can also be drawn from Figure~\ref{fig:ipad_pop}.

\subsection{Data Leakage in Recommender System}
\label{ssec:leakageExplaination}

Data leakage is described as ``the use of information in the model training process which would not be expected to be available at prediction time'' in Wikipedia.\footnote{\url{https://en.wikipedia.org/wiki/Leakage_(machine_learning)}} Deemed ``one of the top ten data mining mistakes'', Kaufman \etal give a critical discussion on data leakage~\cite{dataleakTKDD} and suggest ``time separation'' of train and test data as one methodology for avoiding leakage, similar to the \textit{split-by-timepoint} strategy listed in Table~\ref{tab:dataPartitioning} in our context. Note that data leakage is fundamentally different from leak in privacy. The former refers to information leakage due to incorrect preprocessing of dataset for model training processes, while the latter suggests the leakage of personal information to other parties.

Among the four data partitioning strategies in Table~\ref{tab:dataPartitioning}, \textit{leave-one-out-split}, also known as leave-last-one-out, does observe local timeline of an individual user. Specifically, if a user has a total of $n$ interactions in a dataset, the model learns user preference from this user's first ($n-1$) interactions, then predicts the user's last (or $n$-th) interaction. At first glance, there is no data leakage with this split strategy because the model learns from the past history (\ie $n-1$ interactions) and predicts a future (\ie the $n$-th) interaction of the same user. Next, we explain why data leakage occurs for this split strategy, because of collaborative filtering.

\begin{figure}
	\centering
	  \begin{subfigure}[t]{0.65\columnwidth}
    	\centering
    	\includegraphics[width=\columnwidth]{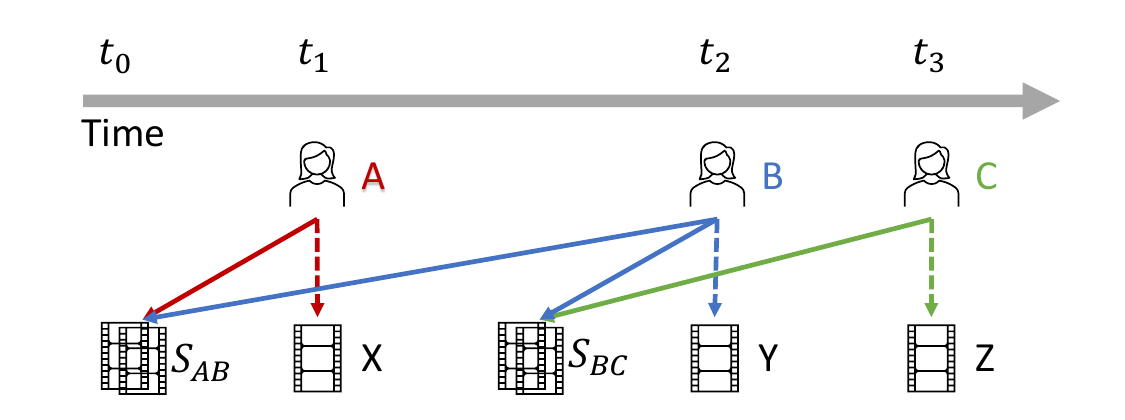}
    	\caption{User-item interaction along global timeline.}
        \label{sfig:userItemTime}
	\Description{}
    \end{subfigure}
	\begin{subfigure}[t]{0.65\columnwidth}
    	\centering
    	\includegraphics[width=\columnwidth]{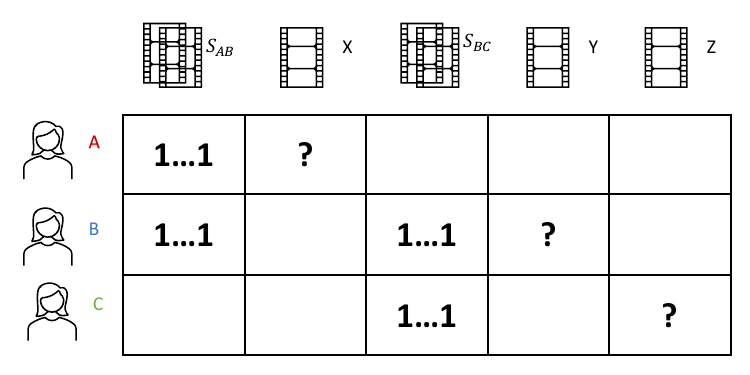}
    	\caption{User-item interaction matrix}
        \label{sfig:userItemMatrix}
	\Description{}
    \end{subfigure}  
	\caption{Interactions along global  timeline, and in matrix form.}
	\label{fig:userItemTimeMatrix}
	\Description{}
\end{figure}

We explain the data leakage problem with an illustration in Figure~\ref{fig:userItemTimeMatrix}. Assuming a system with a global timeline from time $t_0$ to time $t_3$, three example users $A$, $B$, and $C$ interact with items in the system. Figure~\ref{sfig:userItemTime} plots the activities of the users and items along the global timeline. The items are placed in the plot, at the time points they are introduced to the system \ie their release time.\footnote{For easy presentation, we assume that once an item is released for rating, it will stay in the system. That is, we assume an item will not be removed from the system for simplicity.} The users are placed in the plot at the time points of their last interactions. For example, the last rating by user $A$ is on item $X$ at time $t_1$, and we do not observe any more interactions by user $A$ after time $t_1$ in the dataset. $S_{AB}$ denotes the set of items rated by both users $A$ and $B$, and $S_{BC}$ are the items rated by both users $B$ and $C$. 

Assume that we adopt \textit{leave-one-out-split} in this illustration, \ie the last interaction of each user is masked as a test instance. In Figure~\ref{sfig:userItemTime},  test interactions of the 3 users are indicated by dotted lines. Figure~\ref{sfig:userItemMatrix} shows the matrix view of the same example. The test instances are masked with `?', and we use `$1\ldots 1$' to indicate the ratings to multiple items in $S_{AB}$ and $S_{BC}$. 

As aforementioned, without observing global timeline, the entire training set is fed to a recommender as a static set to learn the recommendation model. Because of the common ratings to items in $S_{AB}$, users $A$ and $B$ share high level of similarity; similarly, users $B$ and $C$ share high level of similarity. By learning their latent representations through collaborative filtering, we may find a good level of similarity between users $A$ and $C$. However, recall that all items are plotted in the figure at their release time; all items rated by user $C$ are only available after user $A$'s last interaction \ie time point $t_1$. In reality, these items and their interactions are not possible to be available at time point $t_1$. That is, when predicting user $A$'s preference to item $X$, we are using ``future data'' that are not expected to be available at the prediction time, leading to data leakage. The model may even recommend user $A$ with items that are rated by user $C$ which are ``future items'', because of the learned similarity between users $A$ and $C$.

According to Table~\ref{tab:dataPartitioning}, \textit{random split} of train/test does not observe either local or global timeline. Hence random split leads to data leakage for the same reason as leave-one-out-split.  As these split strategies do not observe local timeline, they may learn from a user's future interactions to predict the same user's current interactions. 

In a recent study~\cite{rigorousEvaluation}, the authors sampled 85 papers on recommender system that were published in top-tier venues (\eg SIGIR, WWW, KDD) in 2017 - 2019. Among the 85 papers, 53.7\% of the papers adopt \textit{random split},  28\% of the papers utilize \textit{leave-one-out}, and 12.3\% of the papers adopt \textit{split-by-timepoint}. That is, only a very small number of offline evaluations follow the ``time separation'' of train and test data to avoid data leakage. The remaining large number of offline evaluations of recommender systems are conducted without considering the issue of data leakage. Hence, most reported results do not well reflect a model's true performance in production setting. 

In this section, we show that the context keeps changing along the global timeline. That is, users may be active or inactive at different time points and items may become available at any time. Hence, it is important to model the temporal context along the global timeline. Failure to consider the global timeline in the offline evaluation of recommender system can lead to data leakage. An immediate question we have is: \textit{what is the impact of data leakage?} The answer to this question is an indication of to what extent the reported performance in literature truly reflects a model's recommendation accuracy in reality.  For instance, if the impact is negligible, then there is no need to change our way of conducting evaluation. Otherwise, we may need to conduct offline evaluation in a more realistic manner for fair comparison of various recommendation models. 
Next, we study the impact of data leakage from the changes in recommendation results with different severity of data leakage, in Sections~\ref{sec:expDesign} and ~\ref{sec:expResults}.

\section{Experiment Design and Setup} 
\label{sec:expDesign}

Evaluating the impact of data leakage is not trivial. Existing offline evaluations adopt multiple train/test split strategies. Among them, only split-by-timepoint does not suffer from data leakage. To avoid the potential influence of random factors, we adopt leave-one-out split in this study. With leave-one-out split, a user's last interaction is masked as a test instance. As shown in Figure~\ref{fig:weekly_activity_with_release_log}, a user's last interaction may occur at any time point along the global timeline. All interactions that occur after a user's last interaction are the ``future data'' to this particular test instance. Here, ``future data'' are interactions from other users. By this definition, the amount of ``future data'' available to each individual test instance is different because test instances can occur at any time along the global timeline. Further, the number of interactions from different users varies significantly. Hence, we argue that studying the impact of data leakage on individual user is computationally expensive. For this reason, we do not study the impact of data leakage at the fine-grained level. 

We design experiments to simulate different severity of data leakage at a coarse-grained level. Briefly speaking, we partition the entire dataset into multiple time windows. Each time window is of the same time interval, \ie one year. Then in the training set, we add user-item interactions in different numbers of time windows to simulate different amounts of ``future data'' available for training. We term this setting as a coarse-grained level, because we do not consider the differences of the future data for individual test instance that happen within the test year.

\begin{figure}[t]
	\centering
    \includegraphics[width=0.75\columnwidth]{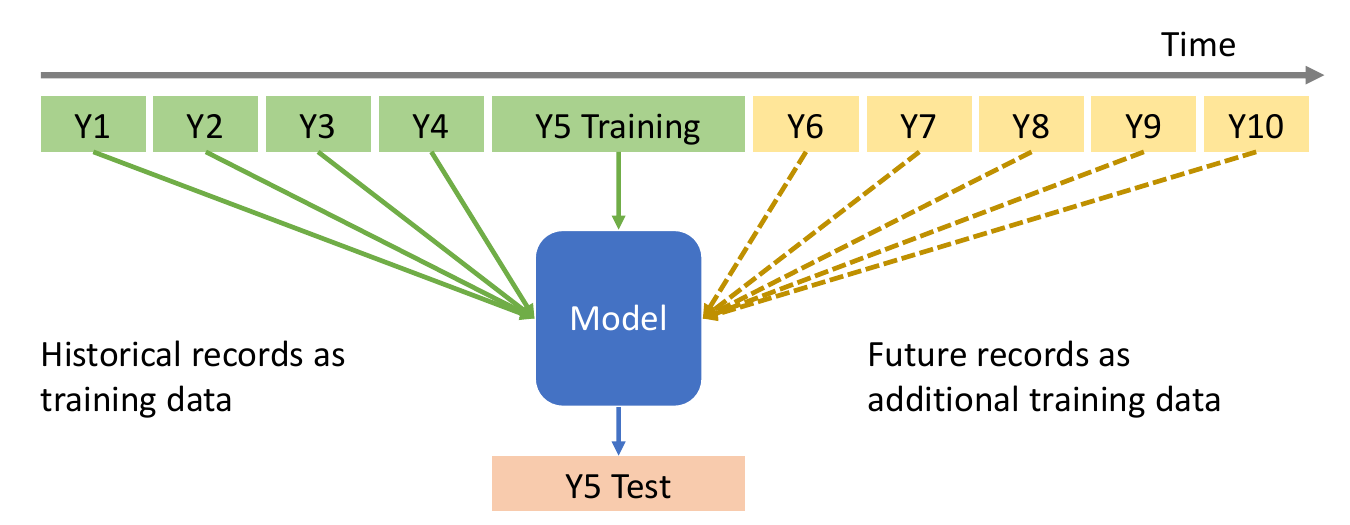}
    \caption{Train/test data split for test year $Y5$ as an example.}
    \label{fig:trainTest}
	\Description{}
\end{figure}

Given a dataset, we first sort all user-item interactions in chronological order and partition the data by a fixed time window. Assume that we have a dataset with a global timeline of 10 years and set the time window to be one year, as shown in Figure~\ref{fig:trainTest}. We refer interactions that occurred in the first year as $Y1$ data. Accordingly, interactions which fall in the last year are referred to as $Y10$. We select all test instances in $Y5$ as the subset for evaluation, \ie the users' last interactions fall within year 5. All interactions that occur before $Y5$, \ie $Y1$ to $Y4$, are ``historical interactions'' with respect to the test set in $Y5$. Correspondingly, we have interactions from $Y6$ to $Y10$ as ``future interactions''. We then conduct several runs of experiments to simulate different severity of data leakage. In the first experiment, the training data consists of all ``historical interactions'' from $Y1$ to $Y4$ and all training instances in $Y5$, highlighted in green in Figure~\ref{fig:trainTest}. Training instances in $Y5$ are the interactions that fall in year 5 and are not any user's last interaction.  The second run's training data will further include all interactions in $Y6$. In the third run, all data in $Y7$ are added as additional training instances. This process continues until all user-item interactions till $Y10$ are used for training. With this setting, we are able to replicate a gradually intensifying data leakage problem under leave-one-out-split setting for test instances occurred in $Y5$. The first run of the experiment has the least data leakage, whereas the last run suffers the worst data leakage. We study the recommendation results of the models  to quantify the impact of data leakage. 

\subsection{Dataset} 
\label{ssec:dataset}

We use MovieLens-25M, Yelp, Amazon-music, and Amazon-electronic datasets in our experiments. We use these four datasets for two reasons. First, they are all widely used datasets in both time-aware recommender system and non-time-aware recommender system~\cite{rigorousEvaluation}. Second, all these datasets contain the timestamps of ratings, so that we can reconstruct the global timeline.

We treat ratings in all four datasets as implicit feedback, and formulate the recommendation problem as \textit{top-$N$} recommendation rather than rating prediction. From each dataset, we extract 10-year interaction data and remove duplicated data. The statistics of the four datasets are summarized in Table~\ref{tab:datasetStats}. As a common practice, we filter inactive users and items with $k$-core filtering on Yelp, Amazon-music, and Amazon-electronic datasets. With $k$-core filtering, datasets are filtered until all users and items have at least $k$ interactions. On Yelp and Amazon-electronic, we apply 10-core filtering. To keep a reasonable number of users and items, we apply 5-core filtering on Amazon-music. No filtering is applied on MovieLens-25M because the dataset is designed to include only users who have at least 20 movie ratings. For each dataset, we only keep ratings from users whose first ratings occurred after the starting day of the 10-year time period (see Table~\ref{tab:datasetStats}). If a user has a rating before the first day of the 10-year period, the user will be excluded from the dataset. This is to ensure we have a full picture of every user in the dataset.

We set the time window for evaluation to be 1 year and gradually add ``future interactions'' year by year. We conduct evaluations on two test years, $Y5$ and  $Y7$ respectively. By selecting two test years for evaluation, we can have a comprehensive understanding, avoiding possible bias in a specific year. Recall that an increasing number of ``future training instances'' are added in the experiment to simulate a worsening data leakage problem, we summarize the number of training and test instances in different experiments for each dataset, in Table~\ref{tab:trainTestYears}.

\begin{table*}
  \begin{center}
    \caption{Number of test and training instances for test years $Y5$ and $Y7$. From the test year till $Y10$, more future data are made available to training.}
    \label{tab:trainTestYears}
    {\small 
    \begin{tabular}{c|c|c|rrrrrr}
      \toprule
      & Test &\#Test &\multicolumn{6}{c}{ \#Training instances accumulated till adding $Yx$'s data}\\    
      Dataset &   year & instances  & $Y5$ & $Y6$ & $Y7$ & $Y8$ & $Y9$ & $Y10$  \\
      \midrule
      MovieLens& $Y5$ & $3,171$ & $2,489,066$ & $3,876,800$ & $5,602,278$ & $7,243,348$ & $8,474,179$ & $9,805,754$  \\
      -25M&$Y7$ & $9,232$& - & - & $5,596,217$ & $7,237,287$ & $8,468,118$ & $9,799,693$\\
      \midrule
      \multirow{2}{*}{Yelp} & $Y5$ & $3,093$ & $878,494$ & $1,280,070$ & $1,723,554$ & $2,203,266$ & $2,702,445$ & $3,124,122$  \\
      &$Y7$ & $7,241$& - & - & $1,719,406$ & $2,199,118$ & $2,698,297$ & $3,119,974$\\
     \midrule
     Amazon& $Y5$ & $829$ & $18,283$ & $38,873$ & $71,227$ & $95,571$ & $108,496$ & $114,004$ \\
      -music&$Y7$ & $2,686$& - & - & $69,370$ & $93,714$ & $106,639$ & $112,147$\\
     \midrule
     Amazon& $Y5$ & $652$ & $234,398$ & $479,507$ & $898,947$ & $1,317,418$ & $1,607,543$ & $1,751,586$  \\
     -electronic &$Y7$ & $8,747$& - & - & $890,852$ & $1,309,323$ & $1,599,448$ & $1,743,491$\\
      \bottomrule
    \end{tabular}
    }
  \end{center}
\end{table*}

\subsection{Models}  Recommender system is a very active research area and a large number of models have been proposed~\cite{recSysHandbook,recSysSurveyZhang}. In this study, we select four widely used baselines.\footnote{Our implementation is available in GitHub~\url{ https://github.com/putatu/dataLeakageRec}}  Each of the four models belongs to a family of recommendation models.

\begin{itemize}
    \item \textbf{BPR}~\cite{BPR} is a machine learning based model. It proposes to learn a matrix factorization model by pairwise ranking loss. We tune the hyperparameters of BPR by continuous random search. We test latent dimension from $8$ to $128$, learning rate from $1\mathrm{e}{-6}$ to $1\mathrm{e}{-1}$ and regularization coefficient from $1\mathrm{e}{-4}$ to $1\mathrm{e}{-1}$.
    \item \textbf{NeuMF}~\cite{He2017} is a deep learning based model. The model is a combination of matrix factorization and multi-layer perceptron layers to learn a user-item interaction function. Optimal hyperparameters are found by continuous random search. We test latent dimension from $8$ to $128$, learning rate from $1\mathrm{e}{-5}$ to $1\mathrm{e}{-1}$, number of negative instances from $1$ to $4$ and regularization coefficient from $0$ to $1\mathrm{e}{-4}$.
    \item \textbf{SASRec}~\cite{SASRec} is a sequence-aware recommendation model. It uses a self-attentive network to model users' historical interactions and to identify relevant items in a user's interaction history for next item prediction. We tune hyperparameters by continuous random search on learning rate from $1\mathrm{e}{-5}$ to $1\mathrm{e}{-2}$, maximum historical interactions from $1$ to $50$, latent dimension from $8$ to $256$ and number of blocks from $1$ to $4$.
    \item \textbf{LightGCN}~\cite{lightgcn} is a graph neural network based model. It designs a graph convolutional network model for recommender system. Hyperparameters are optimized via continuous random search. We test latent dimension from $8$ to $128$, number of layers from $2$ to $4$, regularization coefficient from $1\mathrm{e}{-5}$ to $1\mathrm{e}{-1}$ and learning rate from $1\mathrm{e}{-5}$ to $1\mathrm{e}{-1}$.
\end{itemize}

The hyperparameters of each model are tuned on the validation set which consists of the second last interactions of the testing users, as in many existing literature~\cite{Bai2019, Huang2018, Pasricha2018}. For each baseline model, we search hyperparameters for $50$ times, and obtain the optimal combination of hyperparameters. Due to space limit, we omit the presentation of optimal hyperparameters. All optimal hyper-parameters can be found in our Github.

\subsection{Evaluation with Non-sampled Metrics} 

In offline evaluation, a recommender recommends users with items from a pool of candidate items. The set of candidate items may not be all items available in the system. According to~\cite{candidateSelections}, there are multiple candidate selection strategies including \textit{TestRatings methodology}, \textit{TestItems methodology}, \textit{TrainingItems methodology}, \textit{AllItems methodology} and \textit{One-Plus-Random methodology}. In our evaluation, instead of making recommendation from a subset of available items, we obtain a top-$N$ recommendation list by considering \textit{a total ranking} of all available items that can be observed in training data for evaluation. That is, we do not use sampled metrics. Sampled metrics in evaluation is to randomly sample a small number (\eg 1000) of irrelevant items, then rank the relevant item only among this small sampled set. A recent study shows that sampled metrics can be inconsistent with non-sampled metrics, and may not truly reflect a model's performance~\cite{EvaluationBias}. Because we use non-sampled metrics, the absolute values of our results are much lower than the values reported based on sampled metrics~\cite{explainableReasoning,He2017}. Understandably, it is much more challenging to rank a relevant item among all irrelevant items, compared with to rank among a much smaller sampled set. 

We evaluate accuracy of the  recommendation models by Hit Rate (\textbf{HR}) and  Normalized Discounted Cumulative Gain (\textbf{NDCG}). Specifically, we report HRs and NDCGs of top-20 recommendations.\footnote{Similar observations hold on the results of top-\{5, 10\} recommendations, hence we do not report results of top-\{5,10\} recommendations due to space limit.}

With the above setup, we analyse the impact of data leakage from two perspectives.  First, we take a detailed look at the top-$N$ recommendation list for each test instance, and study the impact of data leakage on these recommendation lists. Second, we study how HR@20 and NDCG@20 change when different amounts of ``future data'' are added for training.

\section{Experiment Results}
\label{sec:expResults}
Through experiments, we aim to quantify the impact of data leakage from two perspectives: (1) impact on the list of top-$N$ recommended items, and (2) impact on recommendation accuracy. In the remaining part of this section, we  present our findings and provide a detailed explanation of the results.

\subsection{Impact on Top-\textit{N} Recommendation List}  
\label{sssec:resultsLeakage_recList}

\begin{finding}
Models evaluated in our experiments do recommend future items, which are only available in the system after the time point of a test instance.
\end{finding}

We explain the reason behind data leakage in Section~\ref{ssec:leakageExplaination}, with the help of Figure~\ref{fig:userItemTimeMatrix}.  Due to collaborative filtering, a recommender may learn similarity between users $A$ and $C$, illustrated in Figure~\ref{fig:userItemTimeMatrix}, then recommends items rated by user $C$ to user $A$. However, all items rated by user $C$ are only available in the system after user $A$'s last interaction, hence these items are ``future items'' with respect to the time point of user $A$'s last interaction, \ie the test instance for user $A$ in leave-last-one-out split setting. Through experiments, we observe that all four models do recommend future items.

With respect to each test instance in our experiments, we obtain a top-$N$ recommendation list from a recommendation model. Then, we count the number of ``future items'' among these $N$ recommended items. Again, ``future items'' are the items that are released after the timestamp of this test instance. We set $N=20$, and present the number of future items in Table~\ref{tab:FutItems}. Each entry represents the total number of ``future items'' recommended among all the recommendation lists obtained for the test set. We report the numbers obtained for both test years $Y5$ and $Y7$. Reflected in Table~\ref{tab:FutItems}, with more ``future data'' included in training from $Y6$ to $Y10$ for test year $Y5$, and from $Y8$ to $Y10$ for test year $Y7$, there is an increasing trend of ``future items'' recommended by all models on all datasets. 

Recommending ``future items'' for a test instance is strong evidence of data leakage, and the current offline evaluation setting is invalid. Following our evaluation, a similar experiment is conducted in~\cite{futureItemRec}. The authors show that both \textit{random-split-by-ratio} and \textit{temporal-split on a user's local timeline } result in the recommendations of ``future items'', which are invalid recommendations. A recommendation model should not have seen the ``future items'' or any interactions related to these ``future items''. Other than that,  recommending ``future items'' will adversely affect recommendation accuracy  because recommending an item that is not yet available (at the time point of test instance) will never lead to a hit. In such cases, the recommendation accuracy obtained from the offline evaluation is an underestimation of the  model's true performance. However, simply filtering future items from the top-$N$ list before computing recommendation accuracy is not an option. Through the presence of ``future items'', we show that the current offline evaluation setting is invalid, and a model shall not learn from the future data that are not available at the time points of the test instances.

\begin{table*}[t]
  \begin{center}
    \caption{Among top-20 recommendations, the total number of future items recommended for test instances in $Y5$ and $Y7$, respectively, by the four models.}
    \label{tab:FutItems}
    {\small
    \begin{tabular}{l|c|cc|cc|cc|cc} 
      \toprule
    \multirow{2}{*}{Model}  & Dataset &\multicolumn{2}{c|}{MovieLens-25M} &\multicolumn{2}{c|}{Yelp} & 
     
    \multicolumn{2}{c|}{Amazon-music} &\multicolumn{2}{c}{Amazon-electronic} \\
     & Test year  & $Y5$ &  $Y7$ &  $Y5$ & $Y7$ & $Y5$  & $Y7$ & $Y5$ &  $Y7$ \\ 

     \midrule
    \multirow{5}{*}{BPR} 
        &$Y5$ & 0 & $-$ & $0$ & $-$ & 0 & $-$ & $0$ & $-$  \\
        &$Y6$	&$0$	&$-$	&$421$	&$-$	&$615$	&$-$	&$79$	&$-$\\
        &$Y7$	&$22$	&$0$	&$829$	&$0$	&$970$	&$0$	&$363$	&$0$\\
        &$Y8$	&$7$	&$11$	&$2,365$	&$504$	&$1,101$	&$651$	&$263$	&$200$\\
        &$Y9$	&$6$	&$88$	&$5,048$	&$287$	&$1,304$	&$1,103$	&$499$	&$1,224$\\
        &$Y10$	&$4$	&$81$	&$1,851$	&$1,598$	&$1,197$	&$1,155$	&$200$	&$583$\\

     \midrule
     \multirow{5}{*}{NeuMF} 
     &$Y5$ & 0 & $-$ & $0$ & $-$ & 0 & $-$ & $0$ & $-$  \\
     &$Y6$	&$3$	&$-$	&$602$	&$-$	&$910$	&$-$	&$28$	&$-$\\
    &$Y7$	&$7$	&$0$	&$1,631$	&$0$	&$1,501$	&$0$	&$1,303$	&$0$\\
    &$Y8$	&$27$	&$31$	&$3,260$	&$130$	&$1,733$	&$878$	&$549$	&$0$\\
    &$Y9$	&$22$	&$6$	&$3,542$	&$1,177$	&$1,491$	&$1,276$	&$729$	&$216$\\
    &$Y10$	&$15$	&$1$	&$5,205$	&$1,791$	&$1,577$	&$1,573$	&$2,655$	&$326$\\

     \midrule
     \multirow{5}{*}{LightGCN} 
     &$Y5$ & 0 & $-$ & $0$ & $-$ & 0 & $-$ & $0$ & $-$  \\
     &$Y6$	&$11$	&$-$	&$369$	&$-$	&$626$	&$-$	&$37$	&$-$\\
    &$Y7$	&$32$	&$0$	&$739$	&$0$	&$1,050$	&$0$	&$148$	&$0$\\
    &$Y8$	&$116$	&$189$	&$1,070$	&$569$	&$998$	&$632$	&$367$	&$220$\\
    &$Y9$	&$22$	&$26$	&$1,257$	&$979$	&$1,036$	&$893$	&$262$	&$430$\\
    &$Y10$	&$15$	&$58$	&$1,103$	&$1,360$	&$1,152$	&$1,029$	&$260$	&$470$\\

     \midrule
     \multirow{5}{*}{SASRec} 
     &$Y5$ & 0 & $-$ & $0$ & $-$ & 0 & $-$ & $0$ & $-$  \\
     &$Y6$	&$315$	&$-$	&$967$	&$-$	&$906$	&$-$	&$216$	&$-$\\
    &$Y7$	&$442$	&$0$	&$3,074$	&$0$	&$1,548$	&$0$	&$625$	&$0$\\
    &$Y8$	&$144$	&$489$	&$2,228$	&$2,666$	&$1,814$	&$1,341$	&$487$	&$1388$\\
    &$Y9$	&$342$	&$403$	&$3,162$	&$2,893$	&$1,982$	&$1,376$	&$20$	&$3,209$\\
    &$Y10$	&$993$	&$386$	&$1,741$	&$3,014$	&$1,980$	&$1,662$	&$12$	&$2,479$\\

     \bottomrule
    \end{tabular}
    }
  \end{center}
\end{table*}

\newpage
\begin{finding}
Top-N recommendation lists vary with increasing amounts of ``future data''.
\end{finding}

In the previous set of experiments, we tune each run and use the corresponding optimal hyperparameters for recommendation. We note that hyperparameters of a model can be confounding factors for recommendation performance. Hence, in this set of experiments, we remove these confounding factors by using fixed hyperparameters.\footnote{We conduct experiments with SASRec and LightGCN.}

Specifically, given a dataset, a baseline model, and a test year (\eg $Y5$ or $Y7$), we conduct multiple runs of experiments, with different amounts of ``future data'', as specified in Table~\ref{tab:trainTestYears}. Instead of tuning hyperparameters for each run, we fix the hyperparameters to the optimal values obtained for the experiment with no data leakage. By fixing the hyperparameters, we now compare the top-$N$ recommendation sets obtained in different experiments to study the impact of data leakage. That is, given a test instance, we calculate the Jaccard similarity between the top-$N$ items recommended in the experiment without data leakage and the top-$N$ items recommended in the experiment with ``future data''. Given two recommendation lists $L_A$ and $L_B$, the Jaccard similarity $J(L_A,L_B)$ is as follows, by treating a list as a set:
\begin{equation*}
    J(L_A,L_B) = \frac{|L_A\cap L_B|}{|L_A \cup L_B|}
\end{equation*}

The lower the similarity score, the more the data leakage affects the training data distribution, thus the more different the recommended items are from the experiment with no data leakage. 

\begin{figure}
    \centering
    \begin{subfigure}[t]{0.5\columnwidth}
    	\centering
    	\captionsetup{justification=centering}
    	\includegraphics[width=\columnwidth, trim=0 9.5cm 0 0, clip=true ]{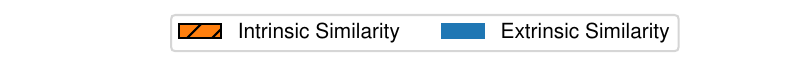}
    	\label{sfig:recList_legend}
	\Description{}
    \end{subfigure}
    \quad
    
    \begin{subfigure}[t]{0.48\columnwidth}
    	\centering
    	\captionsetup{justification=centering}
    	\includegraphics[width=\columnwidth]{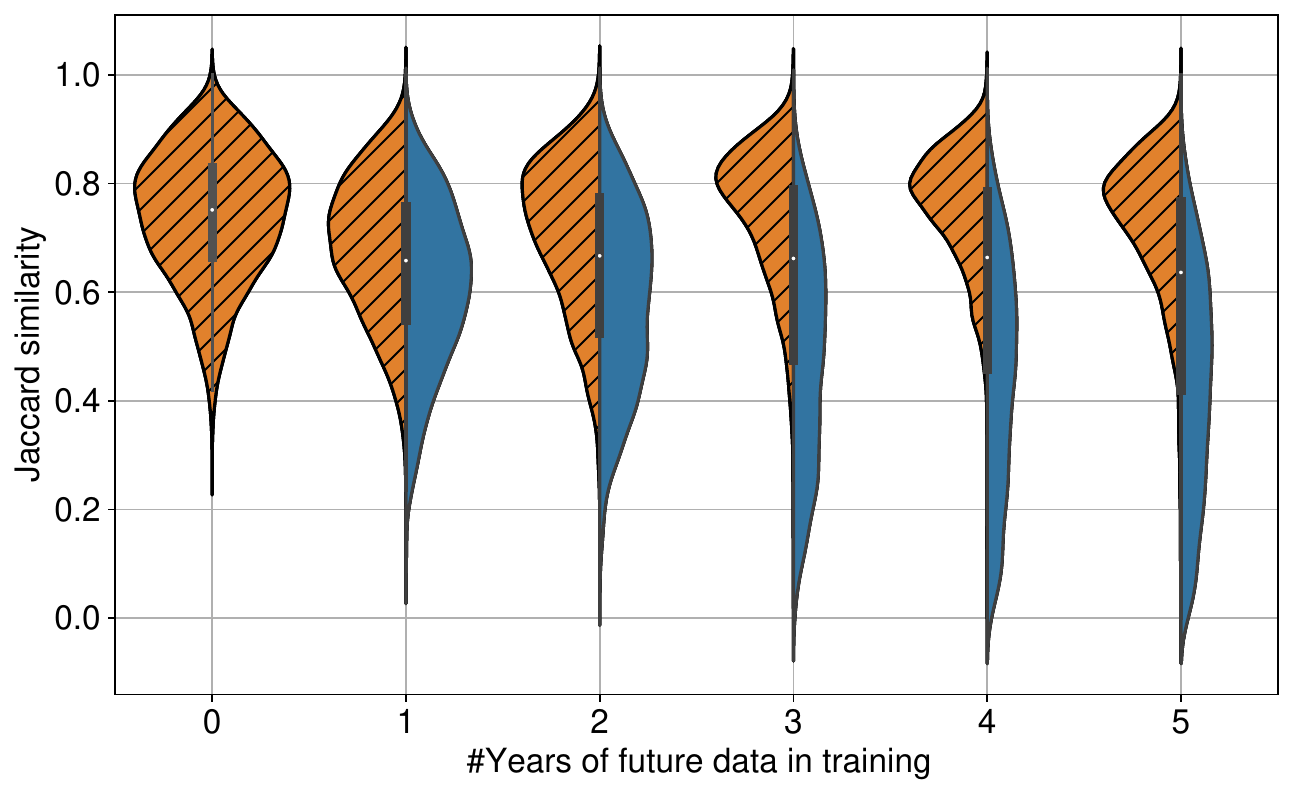}
    	\caption{Y5, LightGCN on MovieLens-25M}
    	\label{sfig:recList_lightgcn_movielens_5}
	\Description{}
    \end{subfigure}
    \quad
    \begin{subfigure}[t]{0.48\columnwidth}
    	\centering
    	\captionsetup{justification=centering}
    	\includegraphics[width=\columnwidth]{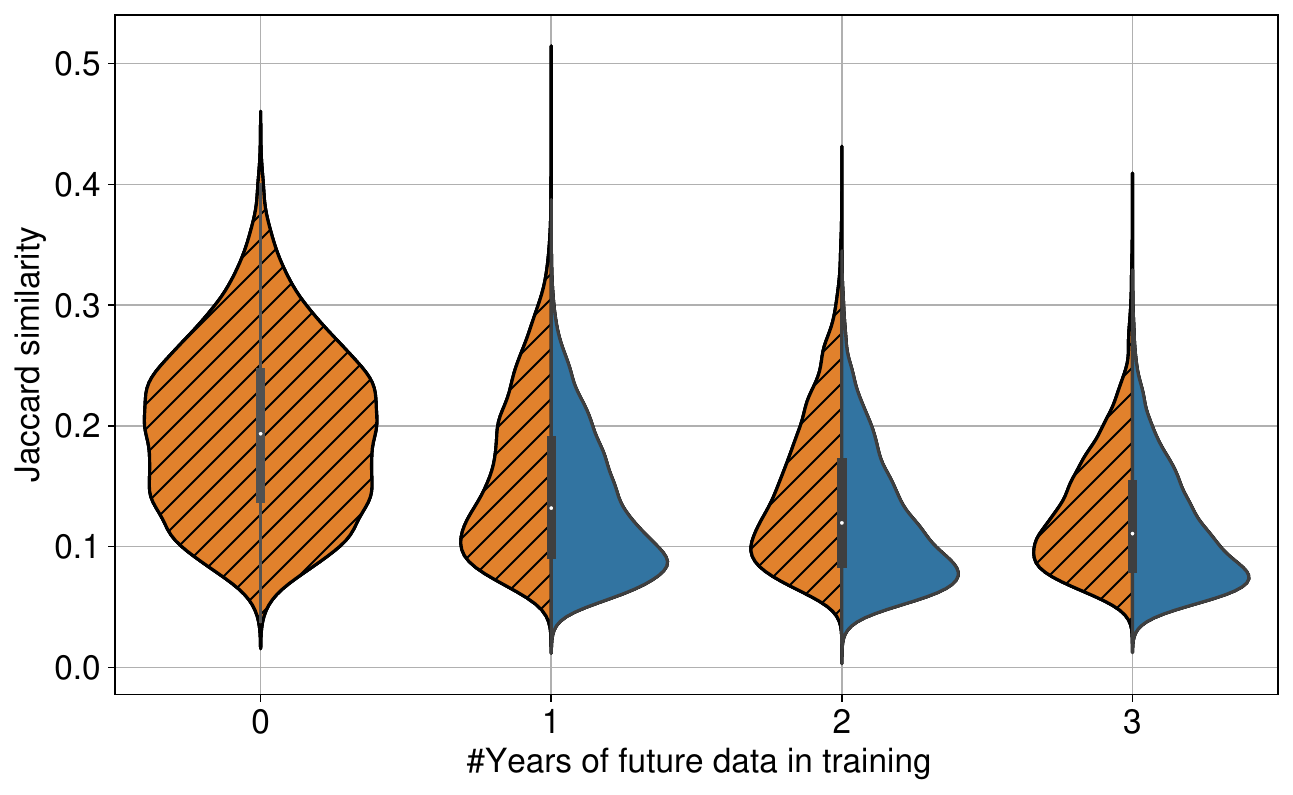}
    	\caption{Y7, SASRec on Amazon-electronic}
    	\label{sfig:recList_sasrec_amazone_7}
	\Description{}
    \end{subfigure}
    \quad
    \caption{Intrinsic Jaccard similarity is calculated from top-$20$ recommendations obtained from the seven runs of experiments on the same training set, but with different seeds. Extrinsic Jaccard similarity is the similarity scores between top-$20$ recommendations from the experiments with no data leakage, and the recommendations with $x$ year of ``future data'' in training.}
    \label{fig:recList}
\end{figure}

For each experiment, we repeat seven random trials with different seeds. Hence, for each test instance in each experiment, we have seven sets of top-$N$ recommendations. We conduct pairwise comparison between the seven sets of recommendations for the same test instance  with no data leakage, and the seven sets obtained with $x$ years of ``future training data''. Then we will have $(7*7) = 49$ similarity values for each test instance. The similarity values are averaged to reduce random error. In Figure~\ref{fig:recList}, we plot the similarity score distributions of all test instances in each experiment.\footnote{Here, we only present comparison results for LightGCN on test year $Y5$ of MovieLens-25M, as well as SASRec on test year $Y7$ of Amazon-electronic. The results for other sets of experiments can be found in our Github.}

We call the similarity obtained from the aforementioned comparisons ``extrinsic similarity'' in Figure~\ref{fig:recList}, because the comparisons are done between experiments with different training data (with and without data leakage). In addition to extrinsic similarity, we further calculate ``intrinsic similarity'' which captures the randomness (due to random seeds) in each experiment with the same training data. Given seven sets of top-$N$ recommendations for each test instance in each experiment, we now have $(7\times 6)/2 = 21$ comparisons, \ie $21$ intrinsic similarity scores for each test instance. The intrinsic similarity values are averaged for each test instance, and the distributions for all test instances are plotted in Figure~\ref{fig:recList} as intrinsic similarity distribution.

Intrinsic similarity sets a reference for extrinsic similarity. The absolute value of extrinsic similarity alone is not very meaningful because there is randomness in the model training process. Hence, we interpret extrinsic similarity by comparing it with the corresponding intrinsic similarity. If the extrinsic similarity distribution is indeed different from its corresponding intrinsic similarity, then the top-$N$ recommendations are indeed affected by the existence of ``future training data''. 

In Figure~\ref{fig:recList}, we plot the results for top-$20$ recommendations, and we made three observations. First, when more ``future data'' is included in training of recommendation models, extrinsic similarity tends to decrease. That is, the top-$20$ recommendation lists become increasingly different from the lists obtained from the experiment without data leakage. This is again a strong evidence that the models are affected by the changed user-item interaction distributions due to the ``future data'' in the training set. The training set with ``future data'' is not a good representation of the past context when the test instance took place. According to Figure~\ref{fig:recList}, MovieLens-25M is more affected by the ``future data'' than Amazon-electronic. A possible reason is that users in MovieLens dataset are only active for very short time periods, and the timestamps in MovieLens are not very reliable. ``Future data'' thus demonstrate a significantly different context from the user's last interactions. Another possible reason is that the recommendation models applied are different on the two datasets. Second, the intrinsic similarity distributions in experiments with ``future data'' are different from the intrinsic similarity distribution in experiment of $0$-year future data. It indicates that the presence of ``future data'' does affect the recommended items for each test instance. Third, extrinsic similarity becomes increasingly different from intrinsic similarity when more ``future data'' is added in training. It indicates that the extrinsic difference exists due to data leakage, but not by randomness in model training processes. 

Recall that the recommendation problem definition in Section~\ref{sec:intro} is learning users' preferences from historical data, then to predict the item a user will rate/interact with in the near future. We show that with ``future data'', the context of ``history'' is not well modeled, hence violating the recommendation problem definition and affecting the recommendation results.

\subsection{Impact on Recommendation Accuracy} 
\label{sssec:resultsLeakage_recPerformance}

\begin{finding}
More ``future data'' can improve or deteriorate recommendation accuracy, making the impact of data leakage \underline{unpredictable}. 
\end{finding}

\begin{figure*}
	\centering
	\begin{subfigure}[t]{0.45\columnwidth}
    	\centering
    	\captionsetup{justification=centering}
    	\includegraphics[width=\columnwidth]{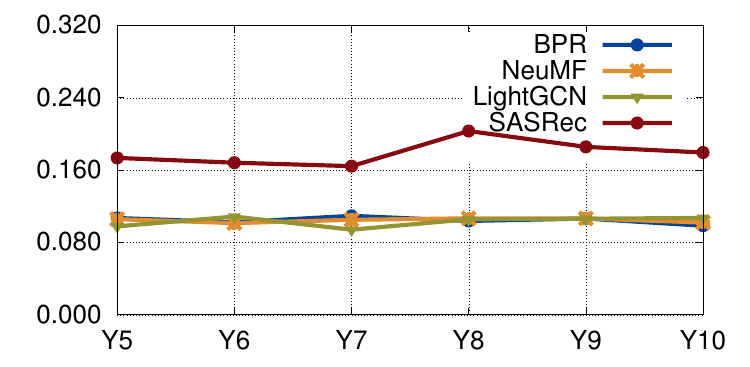}
    	\caption{HR@20 \\MovieLens-25M}
    	\label{sfig:movielens_HR_year5_@20}
	\Description{}
    \end{subfigure}
    \quad
    \begin{subfigure}[t]{0.45\columnwidth}
    	\centering
    	\captionsetup{justification=centering}
    	\includegraphics[width=\columnwidth]{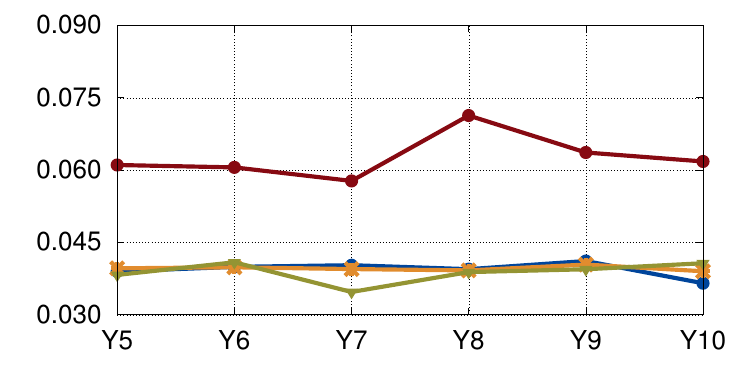}
    	\caption{NDCG@20\\MovieLens-25M}
    	\label{sfig:movielens_NDCG_year5_@20}
	\Description{}
    \end{subfigure}
    \begin{subfigure}[t]{0.45\columnwidth}
    	\centering
    	\captionsetup{justification=centering}
    	\includegraphics[width=\columnwidth]{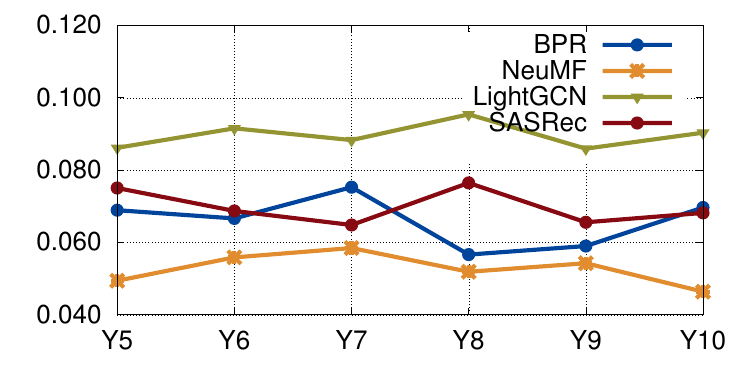}
    	\caption{HR@20 \\Yelp}
    	\label{sfig:yelp_HR_year5_@20}
	\Description{}
    \end{subfigure}
    \quad
    \begin{subfigure}[t]{0.45\columnwidth}
    	\centering
    	\captionsetup{justification=centering}
    	\includegraphics[width=\columnwidth]{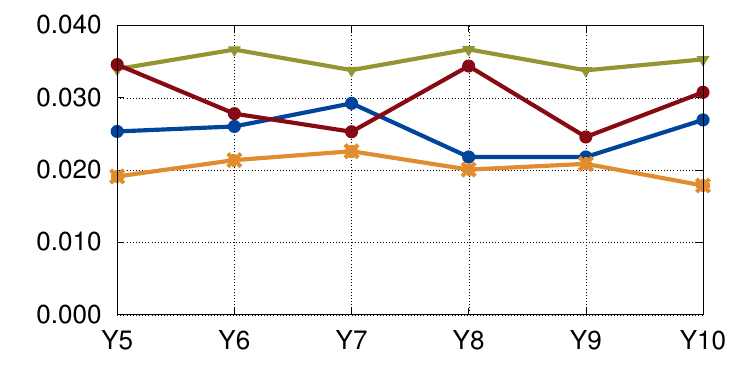}
    	\caption{NDCG@20 \\Yelp}
    	\label{sfig:yelp_NDCG_year5_@20}
	\Description{}
    \end{subfigure}
    \begin{subfigure}[t]{0.45\columnwidth}
    	\centering
    	\captionsetup{justification=centering}
    	\includegraphics[width=\columnwidth]{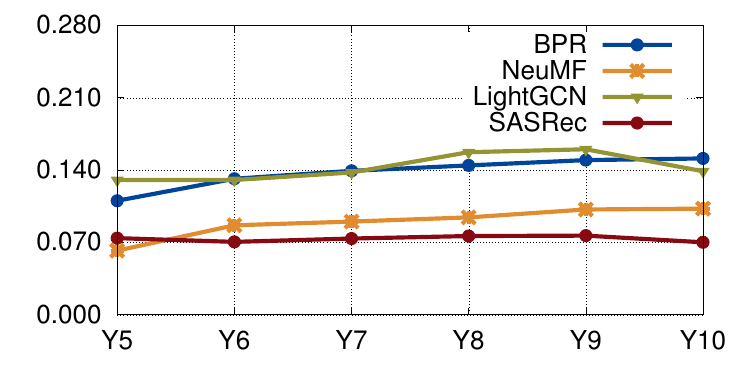}
    	\caption{HR@20 \\Amazon-music}
    	\label{sfig:amazonm_HR_year5_@20}
	\Description{}
    \end{subfigure}
    \quad
    \begin{subfigure}[t]{0.45\columnwidth}
    	\centering
    	\captionsetup{justification=centering}
    	\includegraphics[width=\columnwidth]{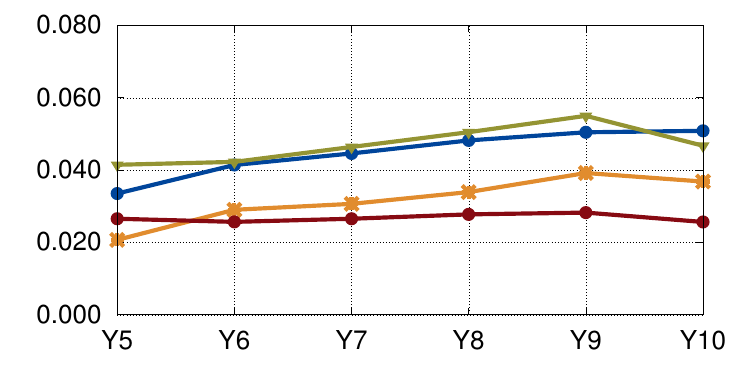}
    	\caption{NDCG@20 \\Amazon-music}
    	\label{sfig:amazonm_NDCG_year5_@20}
	\Description{}
    \end{subfigure}
    \begin{subfigure}[t]{0.45\columnwidth}
    	\centering
    	\captionsetup{justification=centering}
    	\includegraphics[width=\columnwidth]{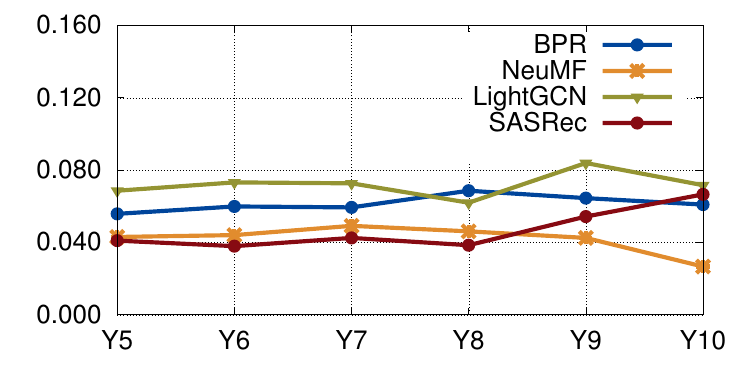}
    	\caption{HR@20 \\Amazon-electronic}
    	\label{sfig:amazone_HR_year5_@20}
	\Description{}
    \end{subfigure}
    \quad
    \begin{subfigure}[t]{0.45\columnwidth}
    	\centering
    	\captionsetup{justification=centering}
    	\includegraphics[width=\columnwidth]{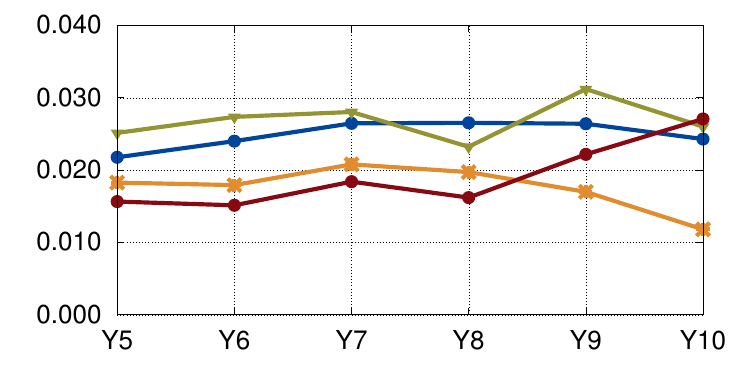}
    	\caption{NDCG@20 \\Amazon-electronic}
    	\label{sfig:amazone_NDCG_year5_@20}
	\Description{}
    \end{subfigure}
    \caption{HR@20 and NDCG@20 for BPR, NeuMF, SASRec and LightGCN on test year $Y5$, along the inclusion of future data from $Y6$ to $Y10$.}
    \label{fig:dataLeakageRecPerformance}
\end{figure*}

Selecting $Y5$ as the test year, in Figure~\ref{fig:dataLeakageRecPerformance}, we plot the HR@20 and NDCG@20 values of BPR, NeuMF, SASRec, and LightGCN, on the four datasets. The HR@20 and NDCG@20 values are the average values of three random trials of the same experiments with different seed numbers. By taking average, we aim to reduce the potential noises from random factors. Recall that we use non-sampled metrics, as explained in Section~\ref{sec:expDesign}, the values of each metric reported in Figure~\ref{fig:dataLeakageRecPerformance} are much lower than the values reported in other papers~\cite{He2017,explainableReasoning} which use sampled metrics. 

Referring to Figure~\ref{fig:dataLeakageRecPerformance}, we observe fluctuating patterns for all four models when more ``future data'' are added in training, from $Y6$ to $Y10$. In particular, ups and downs can be seen for HRs and NDCGs on MovieLens-25M and Yelp datasets. Similarly, fluctuating pattern can be observed for LightGCN on Amazon-music, and for BPR, LightGCN on Amazon-electronic. We also observe decreasing trend in NeuMF on Amazon-electronic. An increasing pattern is seen for BPR, NeuMF and SASRec on Amazon-electronic. We observe not much change in HR@20 and NDCG@20 by SASRec on Amazon-music, the smallest dataset among the four.

In summary, more ``future training instances'' do not necessarily lead to better or worse recommendation accuracy, but do affect values. Hence, recommendation results become unpredictable when there are ``future data'' (with respect to the test instances) in the training set. The same plot on the test year $Y7$ can be found in Appendix~\ref{appendix:expResults}. We note that experiments in test year $Y7$ give results similar to those of test year $Y5$. That is, data leakage leads to unpredictable recommendation performance.

\begin{table*}
  \begin{center}
    \caption{Lowest, highest performance changes (in percentage) when having more future data from the test year till $Y10$, computed based on the result of not accessing future data in training as reference.}
    \label{tab:improvement}
    \begin{tabular}{c|c|c|c|c|c} 
    \toprule
    Dataset&Metric & BPR  &NeuMF & LightGCN  & SASRec\\
    \midrule
    MovieLens&HR@20	&$-8.0\%$, $+2.3\%$	&$-4.1\%$, $+0.9\%$	&$-3.8\%$, $+11.1\%$	&$-5.3\%$, $+17.2\%$ \\
    -25M&NDCG@20	&$-6.3\%$, $+5.5\%$	&$-1.5\%$, $+2.0\%$	&$-9.3\%$, $+6.8\%$	&$-5.4\%$, $+16.8\%$\\  

    \midrule
    \multirow{2}{*}{Yelp}&HR@20	&$-17.8\%$, $+9.2\%$	&$-6.1\%$, $+18.3\%$	&$-0.3\%$, $+10.8\%$	&$-13.6\%$, $+1.9\%$\\
    &NDCG@20	&$-13.9\%$, $+15.4\%$	&$-6.6\%$, $+18.3\%$	&$-0.5\%$, $+8.0\%$	&$-29.0\%$, $-0.6\%$\\
    \midrule
    Amazon&HR@20	&$+19.3\%$, $+37.2\%$	&$+39.6\%$, $+65.6\%$	&$0\%$, $+22.8\%$	&$-5.4\%$, $+3.3\%$\\
    -music&NDCG@20	&$+23.6\%$, $+51.8\%$	&$+40.2\%$, $+89.5\%$	&$+1.9\%$, $+32.7\%$	&$-3.4\%$, $+6.3\%$\\
    \midrule
    Amazon&HR@20	&$+6.4\%$, $+22.9\%$	&$-38.1\%$, $+14.3\%$	&$-9.7\%$, $+22.4\%$	&$-7.5\%$, $+62.5\%$\\
    -electronic&NDCG@20	&$+10.3\%$, $+22.0\%$	&$-35.5\%$, $+13.8\%$	&$-7.7\%$, $+24.1\%$	&$-3.3\%$, $+73.0\%$\\
    \bottomrule
    \end{tabular}
  \end{center}
\end{table*}

We further summarize the unpredictability of recommendation accuracy in Table~\ref{tab:improvement}. In this table, we use the results of not adding any future data in training as the reference results, and report the lowest and highest changes in performance when future data are added till $Y10$. Using the test year $Y5$ and HR@20 as an example, HR@20 on $Y5$ is the reference; we compute the percentage of changes in HR@20's when adding future data from $Y6$ to $Y10$, and report the lowest and highest percentage changes between $Y6$ and $Y10$.\footnote{We only report HR@20 in Table~\ref{tab:improvement}. Other metrics like HR@10 and NDCG@10 give similar results.} Shown in Table~\ref{tab:improvement}, with respect to the results of no data leakage, the performance changes largely in both signs and magnitudes. Specifically, the magnitudes in Table~\ref{tab:improvement} can be as high as $89.5\%$, and there is no particular pattern. In short, the impact of data leakage on recommendation results is unpredictable. Therefore, the recommendation performance reported from experiments with ``future data'' is not reflective of the actual performance in a more practical setting without data leakage.

\begin{finding}
The evaluated models do not show consistent patterns in terms of their relative performance ordering, with the inclusion of additional future data.
\end{finding}

Referring to plots on MovieLens-25M, Yelp, Amazon-music, and Amazon-electronic in Figure~\ref{fig:dataLeakageRecPerformance}, ranking order of the four baselines changes when more ``future data'' are added in training, from $Y6$ till $Y10$. We summarize the ranking orders of the four baselines in Table~\ref{tab:rankingOrder}. Here, $1$ indicates that the model performs the best (\ie with the highest HR@20), while $4$ refers to the worst among the four baseline models. We only present the ranking order of each baseline model for the test year $Y5$ for the sake of space. More details can be found on Github.

Listed in Table~\ref{tab:rankingOrder}, we observe inconsistent ranking orders of the four baseline models when different amount of ``future data'' is added for training. Specifically, on MovieLens-25M, although the sequential recommender SASRec consistently performs the best, there are multiple swaps between the ranking orders of other recommenders BPR, NeuMF, and LightGCN. We cannot decide which general recommender will give better recommendation results. Similarly, on Yelp,  LightGCN performs the best and NeuMF performs the worst, no matter how much ``future data'' is added. However, the relative performance ordering between BPR and SASRec is inconsistent and unpredictable. The inconsistent ranking orders of the four baseline models can also be observed on both Amazon-music and Amazon-electronic datasets. 

Note that all models are tested on the same set of test instances from $Y5$, and the only difference is the inclusion of future data (from $Y6$ to $Y10$) that are not supposed to be available in $Y5$. The inconsistent patterns hinder us from deciding which model would in general give better recommendation results.

\begin{table*}[t]
    \centering
    \caption{Ranking order of BPR, NeuMF, SASRec, and LightGCN in terms of HR@20 for test year $Y5$,  with different amount of ``future data'', on four datasets. Here, $1$ indicates that the model achieves the highest HR@20, and correspondingly the best recommendation performance. $4$ refers to the worst model in terms of HR@20 among four baselines.}
    {\small 
    \begin{tabular}{c|c|cccc}
        \toprule
         Dataset & Train Year & BPR & NeuMF & SASRec & LightGCN \\
         \midrule
         \multirow{6}{*}{MovieLens-25M} &$Y5$	&$2$	&$3$	&$1$	&$4$\\
        &$Y6$	&$3$	&$4$	&$1$	&$2$\\
        &$Y7$	&$2$	&$3$	&$1$	&$4$\\
        &$Y8$	&$4$	&$2$	&$1$	&$3$\\
        &$Y9$	&$3$	&$2$	&$1$	&$4$\\
        &$Y10$	&$4$	&$3$	&$1$	&$2$\\

         \midrule
         \multirow{6}{*}{Yelp}&$Y5$	&$3$	&$4$	&$2$	&$1$\\
        &$Y6$	&$3$	&$4$	&$2$	&$1$\\
        &$Y7$	&$2$	&$4$	&$3$	&$1$\\
        &$Y8$	&$3$	&$4$	&$2$	&$1$\\
        &$Y9$	&$3$	&$4$	&$2$	&$1$\\
        &$Y10$	&$2$	&$4$	&$3$	&$1$\\

         \midrule
         \multirow{6}{*}{Amazon-music}&$Y5$	&$2$	&$4$	&$3$	&$1$\\
        &$Y6$	&$1$	&$3$	&$4$	&$2$\\
        &$Y7$	&$1$	&$3$	&$4$	&$2$\\
        &$Y8$	&$2$	&$3$	&$4$	&$1$\\
        &$Y9$	&$2$	&$3$	&$4$	&$1$\\
        &$Y10$	&$1$	&$3$	&$4$	&$2$\\

         \midrule
         \multirow{6}{*}{Amazon-electronic} &$Y5$	&$2$	&$3$	&$4$	&$1$\\
        &$Y6$	&$2$	&$3$	&$4$	&$1$\\
        &$Y7$	&$2$	&$3$	&$4$	&$1$\\
        &$Y8$	&$1$	&$3$	&$4$	&$2$\\
        &$Y9$	&$2$	&$4$	&$3$	&$1$\\
        &$Y10$	&$3$	&$4$	&$2$	&$1$\\

         \bottomrule
    \end{tabular}
    }
    \label{tab:rankingOrder}
\end{table*}

\subsection{Summary on Impact of Data Leakage} 

In summary, in our experiments, we study the results of the recommendation from two perspectives. One is the top-$N$ recommended items made by the evaluated models. The other is the recommendation accuracy and the relative performance ranking orders among the four models evaluated. 

As a strong evidence of data leakage, we find that ``future items''  present among the top-$N$ recommended items. In reality, these ``future items'' are only available in the system after the timestamps of the test instances. It is impossible to recommend a phone that is released in 2022 to a user at a time in 2020.
Recommending a future item at a past time point is not realistic and it is a sign of invalid offline evaluation setting.  We further show that when more ``future data'' are added to the training set, the top-$N$ recommended items become more dissimilar from the recommendation lists when there is no data leakage. Again, this result supports our argument that user-item interaction distributions change along the addition of ``future interactions''. 

Through experiments, we find that presence of ``future data'' in training  indeed affects recommendation accuracy. We show that HR@20 values differ between experiments with no ``future data'' and experiments with  ``future instances'' in training. Hence, when data leakage exists, the recommendation performance is no longer reflective of a model's actual performance in realistic or production setting.   We also observe the swapping of ranking positions in terms of recommendation performance among the four evaluated models. It shows that the presence of ``future data'' makes it impossible to compare model performance and to decide which model will perform better in recommendation. Our findings are consistent with and explain results in earlier studies~\cite{exploreDataSplit, rigorousEvaluation}, that reproducibility of recommendation performance by baseline models is affected by data splitting strategies. We show that different amount of ``future data'' leads to unpredictable  ranking order, thus impeding the reproducibility of recommendation performance.

Findings from our experiments call for a revisit of the offline evaluation setting, for a better simulation of the actual context when the test instances took place.

\section{Timeline Scheme: Toward a more realistic evaluation}
\label{sec:timelineScheme}

In Section~\ref{sec:intro}, we argue that not observing global timeline leads to two major issues. One is that the designed models are not able to capture the global temporal context of user-item interactions. The other issue is data leakage. Through experiments, we have explored the impact of data leakage with the \textit{leave-one-out} data split and commonly used baseline models. In fact, all data split schemes (\eg random split) that do not observe global timeline would suffer from the same. The fluctuating performance also suggests that the performance comparison reported in literature is hard to reproduce. In this sense, our finding is consistent with the findings reported in~\cite{worrying,exploreDataSplit, rigorousEvaluation} which show inconsistent performances from the recommenders by using different data split schemes. Our analysis further explains their observations from the perspective of ignoring the global timeline. 

\begin{figure}
	\centering
    \includegraphics[width=0.75\columnwidth]{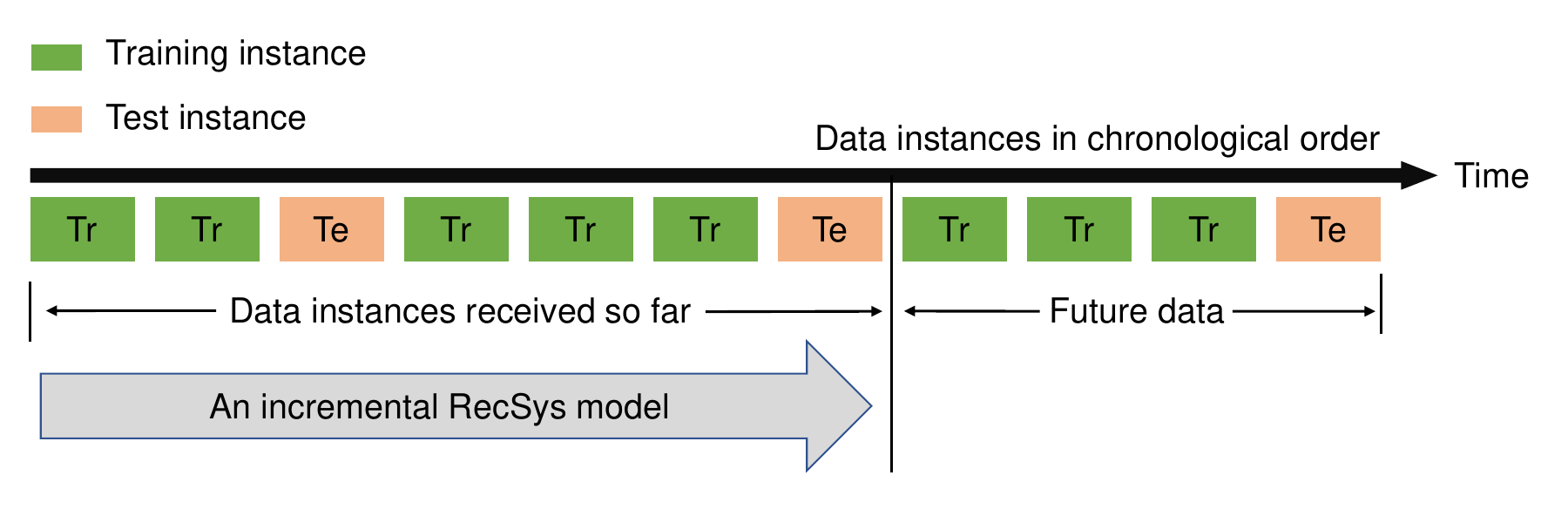}
    \caption{An illustration of the timeline scheme. An incremental recommendation model takes in data instances along time, and produces recommendations when a test instance is encountered. Or, upon encountering a test instance, retraining of a batch-learning model is conducted using historical interactions occurred before the test instance. Test instances are pre-determined by a data split scheme \eg leave-one-out or random split.}
    \label{fig:timeline_scheme}
	\Description{}
\end{figure}

Among the commonly used data partitioning strategies listed in Table~\ref{tab:dataPartitioning}, \textit{split-by-timepoint} does not suffer from  data leakage problem. Despite that,  simply adopting split-by-timepoint may not necessarily lead to more realistic evaluation of recommenders. Split-by-timepoint sets aside a time period for evaluation with a predefined time point. All interactions that occurred before the time point are used as training instances. If the test time period is too short, there may be not sufficient test instances to give a fair and unbiased evaluation result. If the test time period is too long, then the temporal context of user-item interactions during the test period may have changed. For instance, items show different popularity patterns along the timeline in Figure~\ref{fig:pop_globaltimeline}. A long test time period prevents a model from learning these temporal context changes along the global timeline. Again, the evaluation results may not be reflective of how the recommender system performs in a real-time setting. Taking this into account, it would be difficult to decide the duration of the testing period.

We consider a \textit{timeline scheme} to be a more reasonable scheme for fair offline evaluation of recommender models. With the timeline scheme, all user-item interactions are sorted in chronological order, as illustrated in Figure~\ref{fig:timeline_scheme}. A subset of the interactions is masked as test instances. The sampling of test instances can be by random or even by leave-one-out (\eg each user's last interaction is masked as test). A recommender shall take in training instances \textit{along the timeline} to learn/update its model. Whenever a test instance is encountered along the timeline, the recommender makes prediction based on (i) what the model has learned up to that time point, and (ii) the current pool of available items at that time point. This timeline scheme observes global timeline by design. Not all training instances are fed to learn the model as a whole. Instead, the training instances are received along the timeline till the time point when a recommendation is to be made, preventing data leakage. Test instances are sampled along the timeline, hence the model is able to learn the temporal context along timeline and make recommendations accordingly. 

In the extreme case, the timeline scheme becomes prequential evaluation, a commonly used evaluation scheme in data stream mining. In prequential evaluation, all data instances are sorted in chronological order. Each instance is first used to test a model, and then its true label is released to train the model. In fact, prequential evaluation has been adopted in online recommendations~\cite{adaptiveRecsys,onlineRecsys,streamRecsys}. The timeline scheme can be considered as a relaxed version of the prequential evaluation. 

If we relax the timeline scheme, we may use a sliding time window as a test period, \eg all interactions occurred in the past few week are used to learn a model to make predictions for the next week, along the global timeline. In this example, the time window is one week, and the model is trained/updated on a weekly basis with the interactions available before the test week. Adoption of this scheme requires consideration of the following factors: 

\begin{enumerate}
    \item [(1)] \textit{Length of time window}. If the time window is set too long, the recommendation model may encounter severe cold-start issue because many users/items are only active some time after the starting day of global timeline. If the time window is too short, it may be computationally expensive because there would be extensively large number of sliding windows. To decide on the length of time window, data characteristics such as dynamics in the dataset and severity of cold-start problem in the dataset should be analyzed.
    
    \item [(2)] \textit{Hyperparameter tuning}. As multiple evaluations will be conducted using this scheme, hyperparameter tuning strategy becomes an important topic to study. One possible strategy could be tuning using a subset of time windows~\cite{SML, graphsail}. That is, the global timeline can be divided into $n$ time windows. Following the relaxed version of \textit{timeline scheme}, there would be $n-1$ times of evaluations conducted (the first time window will not be tested because the recommendation model has not learned from any data instances). From the $n-1$ evaluation windows, the $m$ time windows are selected as the validation set for hyperparameter tuning, with $m < n-1$. The recommendation accuracy on the $m$ time windows will be used as metrics for the selection of hyperparameters. We note that selection of the $m$ validation time windows also requires careful designs to avoid data leakage. A possible method is to select the $m$ time windows that happen just before the test set (the time windows set aside for reporting of the recommendation accuracy).
    
    \item [(3)] \textit{Interpretation of recommendation accuracy over time}. As evaluations can be done in multiple time windows, multiple sets of recommendation accuracy would be reported. According to~\cite{timeDepEval,CVTT}, recommendation accuracy can vary over time. A recommendation model may appear to be the best performing model at the beginning, but becomes less competitive as time passes. Hence, a research question on the interpretation of the recommendation accuracy over time is raised. Possible strategies include averaging recommendation accuracy over a number of time windows or purely considering the recommendation accuracy on the latest time window.
\end{enumerate}

The aforementioned factors apply in the relaxed version of the timeline scheme with sliding windows. Factors (2) and (3) are also concerned in the proposed strict version of the timeline scheme. Careful studies are required before the adoption of the timeline scheme. An example adoption of the timeline scheme is reported in~\cite{CVTT} recently.

We remark that the timeline scheme requires all recommender models to be incremental in nature. Otherwise, batch based retraining using previous historical interactions is required. The former calls for new model designs, as the model training/updating is different from batch-learning models, where all training instances are fed to a model as a whole. The latter calls for a careful design in retraining strategies. Specifically, one needs to decide on the amount of historical interactions needed for retraining. Having all the historical interactions in retraining results in an unscalable solution because the number of historical interactions increases continuously with time. Moreover, not all historical interactions are useful in learning user preference because preference can change over time~\cite{loyaltyRec}. Nevertheless, solely using a subset of historical interactions may lead to forgetting of a user's long-term preference, which is also not desirable in recommender system.

\section{Related Work}
\label{sec:related}

Recommender systems are hard to evaluate. In~\cite{commonPitfalls}, the authors discuss four issues in evaluating recommender systems. These issues include (i) a recommender could be biased towards highly reachable items; (ii) log data obtained from a platform with already installed recommender system could have different distribution from that of interactions obtained from users with no exposure to any recommender; (iii) high click through rate may not translate to high revenue; and (iv) it is hard to evaluate a system based on revenue, because it is questionable whether the users will purchase the items without the recommender. A recent study~\cite{agreeDisagree} explores the disagreement between false-positive metrics and true-positive metrics in evaluating recommender systems. The authors show that false-positive metrics are affected by popularity biases, but in opposite direction compared to true-positive metrics. Another work~\cite{simpsonParadox} highlights that the datasets used in offline evaluation of recommender system suffer from bias caused by the deployed system. Hence, new recommendation models proposed are evaluated based on whether they can reproduce interactions obtained from the deployed system, but not on whether user's preferences are predicted well. While these studies focus on the challenges in evaluating recommender systems in general, we focus on the issues of not observing the global timeline in offline evaluation, which leads to unrealistic evaluations, hence inconsistent performance comparisons.

For offline evaluation, researchers have not reached an agreement on how the evaluation shall be conducted. For example, given a pre-collected static dataset, decisions of offline evaluation settings rely heavily on the choices of dataset splitting scheme~\cite{evaluateRecSys, offlineEvaluation,rigorousEvaluation}. Reported in~\cite{exploreDataSplit}, the performance of published models vary significantly by using different data partition schemes, making the published results non-comparable. We show that leave-one-out split and any other split that do not observe global timeline would lead to data leakage. More importantly, the dynamic contexts behind user-item interactions cannot be well captured by static models. Said and Bellog{\'{\i}}n~\cite{benchmarkRec} make the same observations of non-comparable recommendation results when using different data splitting strategies in offline experiments. The authors of~\cite{rigorousEvaluation} further explore the differences in recommendation results, when using \textit{random-split-by-ratio} and \textit{time-aware-split-by-ratio}, through controlled experiments on  MovieLens-1M, Yelp, Epinions, and Amazon electronic datasets. Random-split-by-ratio randomly samples test data from the whole dataset; time-aware-split-by-ratio considers global timeline and splits dataset into training and test sets by a fixed timestamp. The authors observe that random-split generally leads to better performance than time-aware-split. Based on our study, we argue that this comparison is not very meaningful because random split has the issue of data leakage, hence the recommendation results obtained are unrealistic. Similarly, the authors in~\cite{temporalContext} show  discrepancy between recommendation results obtained from \textit{random-split-by-user} and \textit{strict-temporal cutoff}. They conclude that recommendation results obtained from offline evaluation setting that does not consider time,  do not reflect the actual performance of a recommender in real life. 

Our work is essentially different from the aforementioned papers. We study the non-reproducible results from the perspectives of data leakage rather than the data splitting strategies. Precisely, we do not compare dataset splitting strategies, but explore whether and how data leakage exists in splitting strategy. We also show that data leakage leads to non-comparable recommendation results. Through analysis of the user/item active time period and temporal dynamics of user-item interactions, we argue that split-by-timepoint may not be a good choice. 

``Time'' factor in offline evaluation of recommender system is also studied in~\cite{predictFutureTaste, timeDepEval, CVTT, evaluateDynamic}. The authors of~\cite{predictFutureTaste} acknowledge that a more realistic setting for evaluation of recommendation model is to strictly follow a global timeline. They show that a stricter experiment setting that follows the global timeline leads to different recommendation results from the less strict experiment setting. The authors of~\cite{evaluateDynamic} highlight that a recommender system constantly evolves with new items and new users entering the system. They propose a \textit{temporal leave-one-out} strategy to handle the dynamic properties of recommender system. \textit{Temporal leave-one-out} setting treats each user independently. For each user, it sorts his/her interactions in chronological order. When sliding through a user's local timeline, recommendation is made upon observing a new interaction. Recommendation is made by considering only historical interactions that happened before the timestamp of the new interaction. \textit{Temporal leave-one-out} is similar to our proposed \textit{timeline scheme}. The key difference is that temporal leave-one-out works on user-level and evaluate every interaction observed while our timeline scheme samples interactions from the entire dataset for evaluation. The ``time'' factor is also considered by the authors of~\cite{timeDepEval, CVTT}. They conduct experiments to study how recommendation performance changes over time with more interactions being observed. It is shown that a recommendation model's performance varies over time and it is not true that a recommender model will consistently outperform the other models at any time. This finding suggests context changes over time in recommendation. Hence, it further confirms that experiment setting with data leakage is not valid, because the training set is not reflecting the context when a past test instance takes place.

We are not the first to note the issue of data leakage, but the first to offer a comprehensive critical study on this issue from the perspective of ``global timeline''. Many researchers have indicated that time-independent data partition strategies may result in leakage of ``future information'' in training~\cite{predictFutureTaste, timeAware, exploreDataSplit, seqHypergraph}. The data leakage problem, using the leave-one-out strategy as an example, is also briefly discussed in~\cite{exploreDataSplit,popularity20, revisitEval, fairOff}. Although the data leakage problem has been raised, there is a lack of systematic study on the impact of data leakage. 

To our best knowledge,  our work is the first systematic study on the impact of data leakage. We note that our study is different from, and in fact not much related to, time-aware recommendation~\cite{timeAware} and sequential recommendation~\cite{sequentialSurvey}. The key focus of most studies in time-aware recommendation and sequential recommendation is \textit{local timeline} specific to users or items. We focus on the issue of ignoring ``global timeline'' in offline evaluation of recommenders in general, which is applicable to all recommender problems, including time-aware and/or sequential recommendations. 

Stream-based recommender models, as its name suggests, generally follow a global timeline. Some studies~\cite{adaptiveRecsys,onlineRecsys, SML, ASMG} adopt prequential evaluation or enhanced version in their experiments. For example, the instances in the initial time period (\eg first three months of a year) can be used to train a model in batch mode, and prequential evaluation is adopted in the remaining time period, on each new instance~\cite{adaptiveRecsys}. That is, after the initial training period, the user-item interactions are fed to the model in chronological order. Given a new interaction, the recommender model first test the recommendation performance on this interaction and report the performance score. Then the new interaction is used to trained or finetune the recommendation model. Prequential evaluation can be on a set of new interactions~\cite{SML, ASMG, graphsail}, as a relaxed version of the former type. Here, a recommender model is finetuned periodically using instances that occur in the specified time period. In~\cite{streamRecsys}, the authors present an overview of stream-based recommender systems and discuss how to conduct statistical tests for robust evaluation of online incremental recommender models. However, a large number of models published in literature are not incremental in nature. In this paper, through experiments, we highlight the issues of ignoring global timeline in training and evaluating these batch-learning models, \ie non-incremental models. 

\section{Conclusion}
\label{sec:conclude}

In this paper, we provide a critical analysis on data leakage in offline evaluation of recommender systems. The key message is to observe the global timeline in offline evaluation. 

We show the temporal dynamics of users and items through their average active time in four major datasets and highlight that users' last interactions may occur at any time point, and items may be released at any time point along the global timeline. Due to the nature of collaborative filtering, if the train/test data split does not observe the global timeline and all training instances are fed to the recommender as a whole, then the model could learn from data instances that are not available at the time point of the test instance. Through carefully designed experiments, we show that models with data leakage do recommend future items which are not available to the system at the time point of a test instance. We also show that  more future data leads to a more different recommendation lists from the model with no data leakage, for the test instances. Based on recommendation accuracy measures, we show that the impact of data leakage is unpredictable, hence the results reported in literature may not truly reflect performance of recommendation models in online setting. Data leakage does affect the relative performance ranking order of the evaluated models. As a result, it is hard to conclude based on the published literature which models are more likely to give better recommendation results in online setting.  

Based on our understanding of the problem definition of recommender system and what we have learned from this critical analysis, we propose the timeline scheme for a more realistic evaluation of recommender system in an offline setting. Again, the key idea is to follow global timeline and a model shall only learn from interactions that are available before a test instance. The test instances shall also spread along the global timeline to enable models to learn from the temporal context changes. Nevertheless, the proposed timeline scheme requires all models to be incremental, \ie able to learn from the increasingly available training instances along the global timeline. Otherwise, retraining of batch-learning recommendation models is required. We note that a large number of proposed models in literature are not capable of incremental learning. Meanwhile, retraining of batch-learning models, \ie taking all training instances as a whole, may not be feasible because it is difficult to decide on the amount of historical interactions needed for retraining. 

\bibliographystyle{ACM-Reference-Format}
\bibliography{reference}
\appendix
\newpage
\input{Appendix}
\end{document}

%% file: Appendix.tex
\section{Additional Experiment Results}
\label{appendix:expResults}

\subsection{Impact of Data Leakage on Recommendation Performance of Test Year \textit{Y7}}

\begin{figure*}[h]
	\centering
	\begin{subfigure}[t]{0.45\columnwidth}
    	\centering
    	\captionsetup{justification=centering}
    	\includegraphics[width=\columnwidth]{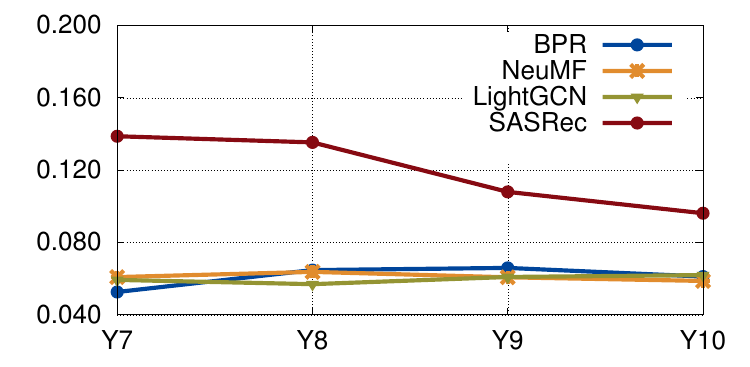}
    	\caption{HR@20, MovieLens-25M}
    	\label{sfig:movielens_HR_year7_@20}
	\Description{}
    \end{subfigure}
    \quad
    \begin{subfigure}[t]{0.45\columnwidth}
    	\centering
    	\captionsetup{justification=centering}
    	\includegraphics[width=\columnwidth]{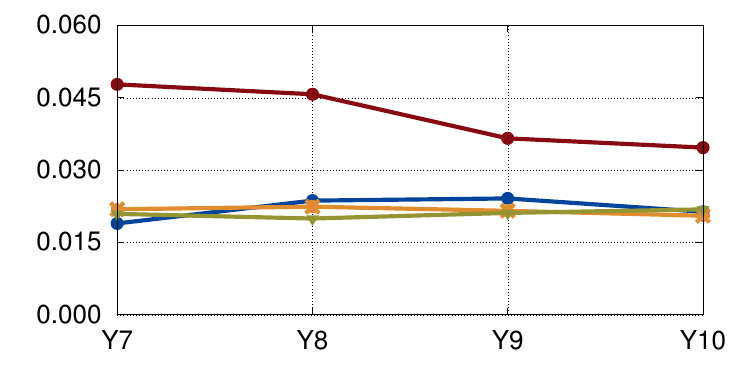}
    	\caption{NDCG@20,MovieLens-25M}
    	\label{sfig:movielens_NDCG_year7_@20}
	\Description{}
    \end{subfigure}
    \begin{subfigure}[t]{0.45\columnwidth}
    	\centering
    	\captionsetup{justification=centering}
    	\includegraphics[width=\columnwidth]{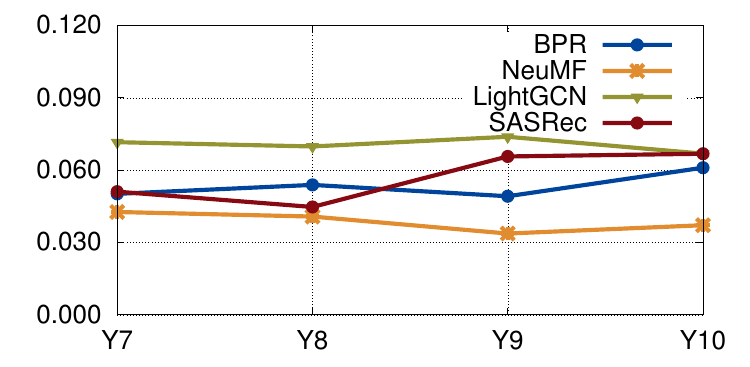}
    	\caption{HR@20, Yelp}
    	\label{sfig:yelp_HR_year7_@20}
	\Description{}
    \end{subfigure}
    \quad
    \begin{subfigure}[t]{0.45\columnwidth}
    	\centering
    	\captionsetup{justification=centering}
    	\includegraphics[width=\columnwidth]{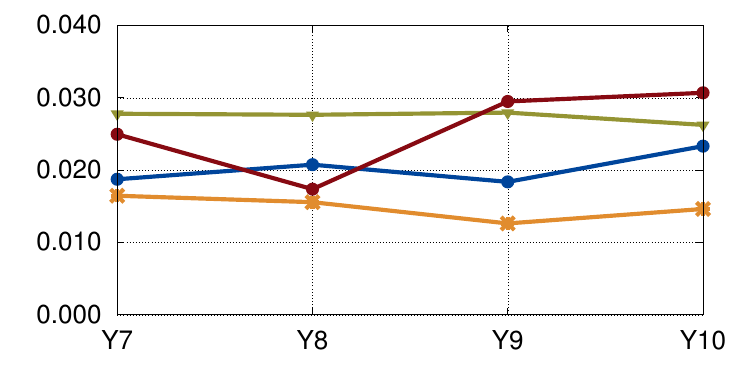}
    	\caption{NDCG@20, Yelp}
    	\label{sfig:yelp_NDCG_year7_@20}
	\Description{}
    \end{subfigure}
    \begin{subfigure}[t]{0.45\columnwidth}
    	\centering
    	\captionsetup{justification=centering}
    	\includegraphics[width=\columnwidth]{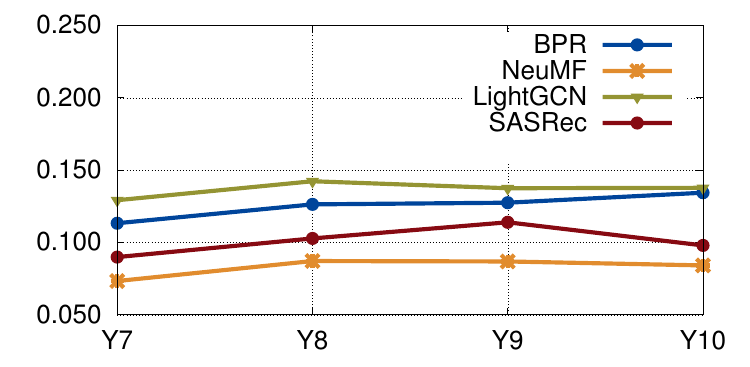}
    	\caption{HR@20, Amazon-music}
    	\label{sfig:amazonm_HR_year7_@20}
	\Description{}
    \end{subfigure}
    \quad
    \begin{subfigure}[t]{0.45\columnwidth}
    	\centering
    	\captionsetup{justification=centering}
    	\includegraphics[width=\columnwidth]{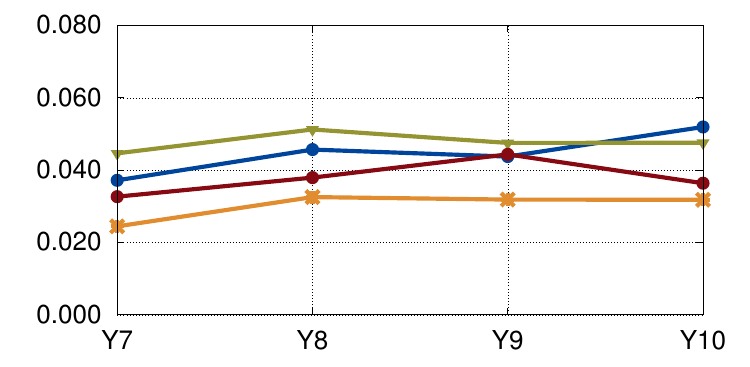}
    	\caption{NDCG@20, Amazon-music}
    	\label{sfig:amazonm_NDCG_year7_@20}
	\Description{}
    \end{subfigure}
    \begin{subfigure}[t]{0.45\columnwidth}
    	\centering
    	\captionsetup{justification=centering}
    	\includegraphics[width=\columnwidth]{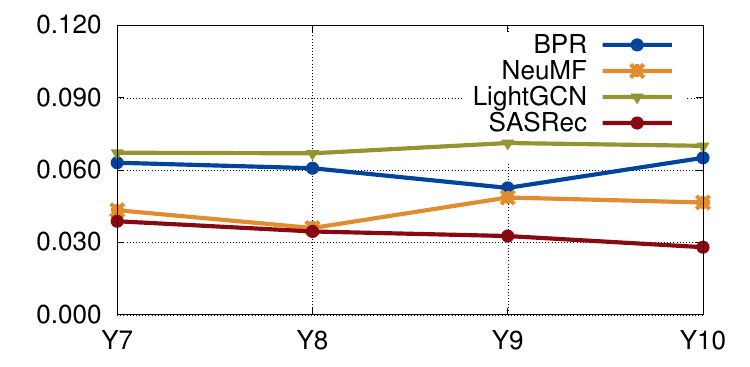}
    	\caption{HR@20, Amazon-electronic}
    	\label{sfig:amazone_HR_year7_@20}
	\Description{}
    \end{subfigure}
    \quad
    \begin{subfigure}[t]{0.45\columnwidth}
    	\centering
    	\captionsetup{justification=centering}
    	\includegraphics[width=\columnwidth]{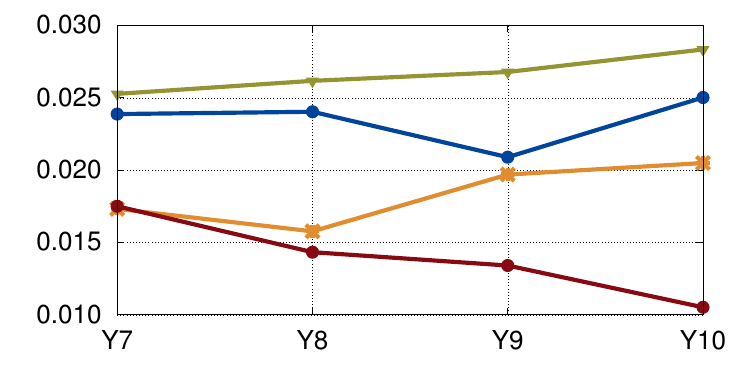}
    	\caption{NDCG@20, Amazon-electronic}
    	\label{sfig:amazone_NDCG_year7_@20}
	\Description{}
    \end{subfigure}
    \caption{HR@20 and NDCG@20 for BPR, NeuMF, SASRec and LightGCN on test years $Y7$.}
\end{figure*}